\documentclass[12pt]{article}
\pdfoutput=1
\usepackage{graphicx}
\usepackage{epstopdf}
\usepackage{amsmath}
\usepackage{amsfonts}
\usepackage{amssymb}
\usepackage{color}
\usepackage{mathrsfs}

\pdfoutput=1

\newcommand{\be}{\begin{equation}}
\newcommand{\ee}{\end{equation}}
\newcommand{\bea}{\begin{eqnarray}}
\newcommand{\eea}{\end{eqnarray}}
\newcommand{\barr}{\begin{array}}
\newcommand{\earr}{\end{array}}

\usepackage[colorlinks,bookmarks]{hyperref}
\definecolor{linkblue}{rgb}{0,0,0.8}
\definecolor{linkgreen}{rgb}{0,0.5,0}

\hypersetup{pdfpagemode=UseNone, pdfstartview=FitH, linkcolor=linkblue, %
            citecolor=linkgreen, urlcolor=linkblue}

\def\beq{\begin{equation}}
\def\eeq{\end{equation}}
\def\be{\begin{equation}}
\def\ee{\end{equation}}
\def\bea{\begin{eqnarray}}
\def\eea{\end{eqnarray}}
\def\d{{\partial}}

\def\nn{\nonumber}

\def\knl{{k_{\rm NL}}}
\def\co{c_{s  (1)}^2}
\def\ct{c_{s  (2)}^2}

\newcommand{\kren}{k_\text{ren}}
\newcommand{\ktr}{k_{\rm tr}}
 \newcommand{\tknl}{{\tilde k}_{\rm NL}}
\newcommand{\invMpc}{\,h\, {\rm Mpc}^{-1}\,}
\def\H{{\cal H}}

\newcommand{\lp}{\left(}
\newcommand{\rp}{\right)}

\begin{document}


\setcounter{page}{1} \baselineskip=15.5pt \thispagestyle{empty}

\begin{flushright}
\end{flushright}

\begin{center}

{\Large \bf The IR-resummed\\[0.5cm] Effective Field Theory of Large Scale Structures}
\\[0.7cm]
{\large Leonardo Senatore${}^{1,2}$  and Matias Zaldarriaga${}^3$}
\\[0.7cm]
{\normalsize { \sl $^{1}$ Stanford Institute for Theoretical Physics,\\ Stanford University, Stanford, CA 94306}}\\
\vspace{.3cm}

{\normalsize { \sl $^{2}$ Kavli Institute for Particle Astrophysics and Cosmology, \\
Stanford University and SLAC, Menlo Park, CA 94025}}\\
\vspace{.3cm}

{\normalsize { \sl $^{3}$ School of Natural Sciences, Institute for Advanced Study, \\Olden Lane, 
Princeton, NJ 08540, USA}}\\
\vspace{.3cm}

\end{center}

\vspace{.8cm}

\hrule \vspace{0.3cm}
{\small  \noindent \textbf{Abstract} \\[0.3cm]
\noindent
We present a new method to resum the effect of large scale motions in the Effective Field Theory of Large Scale Structures. Because the linear power spectrum in $\Lambda$CDM is not scale free the effects of the large scale flows are enhanced. Although previous EFT calculations of the equal-time density power spectrum at one and two loops showed a remarkable agreement with numerical results, they also showed a 2\% residual which appeared related to the BAO oscillations. We show that this was indeed the case, explain the physical origin and show how a Lagrangian based calculation removes this differences. We propose a simple method to upgrade existing Eulerian calculations to effectively make them Lagrangian and compare the new results with existing fits to numerical simulations. Our new two-loop results agrees with numerical results up to $k\sim 0.6 \invMpc$ to within 1\,\% with no oscillatory residuals. We also compute power spectra involving momentum which is significantly more affected by the large scale flows. We show how keeping track of these velocities  significantly enhances the UV reach of the momentum power spectrum in addition to removing the BAO related residuals. We compute predictions for the real space correlation function around the BAO scale and investigate its sensitivity to the EFT parameters and the details of the resummation technique. 

 \vspace{0.3cm}
\hrule


\section{Introduction}

Perturbation theory for LSS dates back to the very early days of modern Cosmology {\it e.g.} \cite{Zeldovich:1969sb,Peebles:LSS}. It is extremely successful at calculating correlators at the lowest order or tree level (for a complete review of perturbation theory results see \cite{Bernardeau:2001qr}). On the other hand, results for the first nontrivial correction to tree level results, the ``loop corrections", are less than satisfactory. These corrections are relevant for upcoming observations and they are not under theoretical control. 

The reason for the failure at the loop level is clear. Perturbation theory cannot be used to describe the small scales because the series simply does not converge in that regime so no resummation of diagrams will fix the problem. In loop calculations, those small scales affect large scale observables as the loop integral cover all momenta. Thus the errors in the small scales pollute large scale results.

This led to the development of the Effective Theory of Large Scale Structure \cite{Baumann:2010tm,Carrasco:2012cv}. This framework explicitly keeps track of the effects of the small scales using a generalized fluid-like description, where the uncertainties produced by the short distance dynamics are encoded  in a set of coefficients which, from the point of view of perturbation theory, are free parameters to be fitted to either simulations or observations. The EFT technique allows one to consistently and systematically keep track of the uncertainties produced by the small scales dynamics that lies outside of the regime of applicability of perturbation theory.

The EFT was originally developed in Eulerian space and used to compute one and two-loop corrections to the matter power spectrum~\cite{Carrasco:2012cv,Carrasco:2013mua} and to study the divergencies that appear in power law Universes~\cite{Pajer:2013jj,Carrasco:2013sva}.

Irrespective of these development, with the renewed interest in modeling the Baryon Acoustic oscillations (BAO),  it has become apparent that, for $\Lambda$CDM, perturbation theory in Lagrangian space is significantly better than its Eulerian counterpart.  Extremely impressive results have been obtained using Standard Perturbation Theory (SPT) both in real and redshift space and also for halos {\it eg.} \cite{Carlson:2012bu}. Furthermore even around the non-linear scale the cross correlation coefficient between the results of an N-body simulation and those of perturbation theory are remarkably better when doing Lagrangian Perturbation Theory (LPT) \cite{Tassev:2011ac,Tassev:2012cq}.  LPT has the same shortcomings when calculation loops as SPT, which motivated us to write the EFT in Lagrangian space in~\cite{Porto:2013qua}.

The difference between Lagrangian and Eulerian perturbation theory can be traced to the fact that there are several different parameters that control the size of non-linearities. In $\Lambda$CDM cosmologies, which have a nontrivial transfer function, these various effects have very different sizes. Thus it is not fully satisfactory to organize perturbation theory in powers of the power spectrum. 

To discuss the various non-linear terms it is convenient to inspect the SPT results for the one loop power spectrum in a Einstein de-Sitter cosmology: 
\be
P= P_{11}+P_{22}+  P_{13}\,,
\ee
with $P_{11}$ being the linear power spectrum and 
\bea
\label{SPToneloop}
P_{13}(k)&=&\frac{a^{4}}{ 252}\frac{k^3}{ 4\pi^2}P_{11}(k)\int dr\,P_{11}(k\,r)\nonumber\\
&&\quad\left({12\over r^2}-158+100r^2-42r^4+{3\over r^3}(r^2-1)^3(7r^2+2)\ln\left|{1+r\over1-r}\right|\right)\label{P13}\,,\\
P_{22}(k)&=&{a^{4}\over 98}{k^3\over 4\pi^2}\int dr\int dx\,P_{11}(k\,r)P_{11}(k\sqrt{1+r^2-2rx})    {(3r+7x-10rx^2)^2\over(1+r^2-2rx)^2}\label{P22}\ .
\eea

One can use (\ref{SPToneloop}) to determine how a mode of wavenumber $k'$ affects the power at the observed wavelength $k$. In the limit $k'\ll k$ one has 
\bea
P_{22}(k)+ P_{13}(k) &\propto& P_{11}(k) \epsilon_{\delta <}\ , \nonumber \\
\epsilon_{\delta <} &=& \int^k_0 {d^3k' \over (2 \pi)^3} P_{11}(k')\ .
\eea
On the other hand for modes $k' \gg k$ the dominant effect comes from $P_{13}$, that scales as:
\bea
P_{13}(k) &\propto& P_{11}(k) \epsilon_{s >} \nonumber \\
\epsilon_{s >} &=&k^2  \int_k^\infty {d^3k' \over (2 \pi)^3}  {P_{11}(k') \over k'^2}\ .
\eea
That is to say, non-linear corrections depend on the variance of the density fluctuations produced by modes with $k'<k$ ($\epsilon_{\delta <}$) and depend on the displacements produced by modes with $k'>k$ ($\delta s_>$) through $\epsilon_{s>}= (k\, \delta s_>)^2$. The fact that modes larger and smaller than $k$ affect the power spectrum through different parameters is what allows SPT to be non-divergent for equal time correlators power law Universes in the range $-3<n<-1$. In this range both  $\epsilon_{\delta <}$ and $\epsilon_{s>}$ are finite. Of course the fact that the result is finite does not guarantee that it is converging to the correct result.
 
The displacements produced by modes with $k'<k$ ($\delta s_<$) do not appear in the equal time correlators we have discussed. They do however  change the final location of those small scale modes and thus can significantly affect some statistics through the parameter $\epsilon_{s<}= (k\, \delta s_<)^2$
\bea
\epsilon_{s_<} &=&k^2  \int_0^k {d^3k' \over (2 \pi)^3}  {P_{11}(k') \over k'^2}\ .
\eea
In fact both $P_{13}$ and $P_{22}$ are directly proportional to $\epsilon_{s<}$,
\bea\label{eq:p22_approx}
P_{22}(k) &\sim& {2\over 3} P_{11}(k) \epsilon_{s <} + \ldots\ , \nonumber \\
P_{13}(k) &\sim& -{2\over 3} P_{11}(k) \epsilon_{s <} + \ldots \ ,
\eea
but $\epsilon_{s <}$ cancels in the final answer,~{ which must be so for general reasons ultimately based on general relativity~\cite{Carrasco:2013sva,Scoccimarro:1995if}.} This is not true for unequal time correlators as in that case $P_{22}$ and $P_{13}$ have a different time dependence. 
What is even more interesting is that in fact it is basically $\epsilon_{s<}$ that is responsible  the broadening of the acoustic peak that degrades the BAO technique even when we are considering equal time correlators. This is so because in $\Lambda$CDM the BAO peak appear as $k$-space oscillations in  $P_{11}(k)$ and the derivation of $P_{22}$ in (\ref{eq:p22_approx}) is inaccurate to treat oscillatory features~\footnote{ The fact that IR-displacements are responsible for the BAO broadening is already quite well known  (see for example~\cite{Eisenstein:2006nk,Crocce:2007dt,Taruya:2012ut,Carlson:2012bu}), but as we describe later, a satisfactory treatment in perturbation theory is not yet available.}. We will discuss that in what follows.

Figure \ref{epsilons} shows the sizes of these $\epsilon$-parameters. Given that the EFTofLSS is expanding in these parameters, convergence can be achieve only where these parameters less than order one~\footnote{ Notice that there are order one ambiguities in the definition of these parameters. It is therefore impossible, without a precise calculation, to determine precisely the convergence radius of perturbation theory.}. It is also clear that  to achieve a desired accuracy one needs to keep more orders in some of these parameters than in others.  The biggest of the parameters are those related to displacements which are dominated by large scale modes and thus are very amenable to perturbation theory. LPT does not expand in  $\epsilon_{s<}$ which in our Universe controls the largest non-linearity in the range of scales of interest for the BAO. We will show in this paper that it is crucial to keep very high orders in $\epsilon_{s<}$ in order to obtain the desired accuracy in  $\Lambda$CDM cosmologies. Thus we will provide formulas where one is not expanding $\epsilon_{s<}$.

\begin{figure}[h!]
\includegraphics[width=0.8\textwidth]{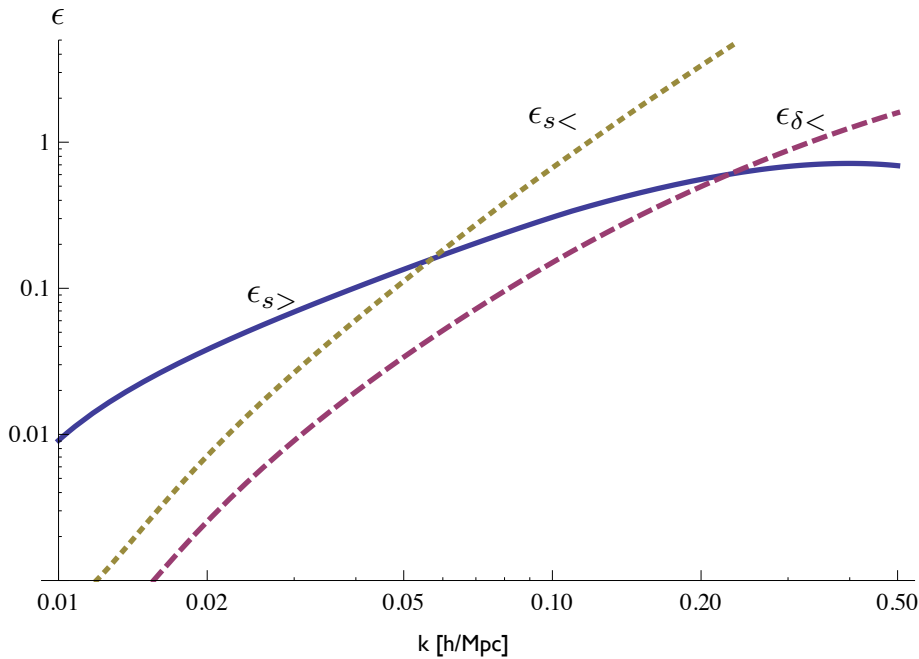}
\caption{\small  Parameters measuring the amplitude of non-linear correction on a mode of wavenumber~$k$. They quantify the motions created by modes longer ($\epsilon_{s<}$) and shorter ($\epsilon_{s>}$) than $k$ and the tides from larger scales ($\epsilon_{\delta <}$). }
\label{epsilons}
\end{figure}

Given the importance of resumming corrections of size $\epsilon_{s<}$ and that these terms will be relevant to get an accurate power spectrum even around the non-linear scale it is useful to gain additional intuition by studying a simple toy model. We  will consider a case, analog to CMB lensing, in which the density field is a Gaussian random field $\delta_L$ which is shifted by a displacement field $\psi$ which is also a Gaussian. The field $\delta_L$ has power spectrum $P_L(k)$ and the field $\psi$ has power spectrum $P_\psi(k)$. For simplicity, we will work in one spatial dimension and take $\delta$ and $\psi$ as uncorrelated. This is a good model to understand the issues as the displacements are dominated by long modes and thus are approximately uncorrelated with fluctuations on the small scales. What the toy model is lacking in the dynamical effects from the long modes, the effects of the tides. 

The model is then
\bea
\delta(x)&=&\delta_L(x+\psi(x)) \nonumber \\
&=& \int {dk \over 2\pi} \hat\delta_L(k) e^{- i k (x + \psi(x))}\ .
\eea
In this toy model the correlation function can be computed exactly
\bea\label{zeta1} 
\xi(x)&=&\langle \delta(x)\delta(0)\rangle = \int {dk \over 2\pi} P_L(k) e^{i k x} \langle e^{i k (\psi(x)-\psi(0))} \rangle \nonumber \\
&=& \int {dk \over 2\pi} P_L(k) e^{i k x}  e^{- k^2 \Delta_\psi(x)/2}\ ,
\eea
where we have defined $\Delta_\psi(x)= \langle(\psi(x)-\psi(0))^2\rangle$. Note that 
\bea
\Delta_\psi(x)&=& 2(\langle(\psi(0))^2\rangle - \langle\psi(x)\psi(0)\rangle) \nonumber \\
&=&  \int {dk \over 2\pi} P_\psi(k) (1-e^{i k x})\ ,
\eea
thus modes with $k x \ll 1$ do not contribute. They shift both points in the correlation function by the same amount.

An important point is that, in equation (\ref{zeta1}),  the contribution to the correlation function  at a separation $x$ coming from modes of wavenumber $k$ is suppressed by $\Delta_\psi$, which receives contributions from all the modes with momentum larger than $1/x$, even those that have momentum smaller than $k$. In other words, in order to contribute, the $\psi$-modes only need to be in the UV with respect to $1/x$, not with respect to $k$. Of course when $\xi(x)$ is featureless, the Fourier transform receives contributions only from modes $k$ up to order $1/x$. But in $\Lambda$CDM $\xi(x)$ is not featureless, as it presents the BAO peak, so that $\xi(x)$ receives contributions also from $k\gg 1/x$.

In this toy model, the loop expansion corresponds to expanding the exponential $\exp{[- k^2 \Delta_\psi(x)/2]}$ in the expression of the correlation function in powers of $\Delta_\psi(x)$. At one-loop we get
\bea\label{zeta1loop} 
\xi^{\rm 1-loop}(x)&=& - \int {dk \over 2\pi} P_L(k) e^{i k x}  k^2 \frac{\Delta_\psi(x)}{2} \nonumber \\
&=& \frac{\Delta_\psi(x)}{2}  \xi''(x)  \nonumber \\
&=& \langle(\psi(0))^2\rangle \xi''(x) - \langle\psi(x)\psi(0)\rangle \xi''(x) \nonumber \\
&\equiv& \xi_{13}(x) + \xi_{22}(x)\ ,
\eea
where we have called the two contributions $\xi_{13}$ and  $\xi_{22}$ because they come from what in the power spectrum we would usually call the $13$ and $22$ terms. Again, notice that the IR cancelation is only for modes that are IR with respect to $1/x$. This runs contrary to the standard intuition:  that is that in the power spectrum at mode $k$, the IR cancellation happens for modes that are long compared to $k$ and not $1/x$.

These simple expressions for the correlation function allow us to estimate the corresponding size of the corrections in perturbation theory.
The correction we have computed in this toy model, that only includes the effects of the displacements and not the dynamical effects, is
\bea
\Delta\xi^{\rm 1-loop}(x) &=& {1 \over 2 } \xi''(x) \Delta_\psi(x)\nonumber \\
&=& {1 \over 2 }  x^2 \xi''(x) {\Delta_\psi(x) \over x^2}\ .
\eea 
In our Universe, this term is very enhanced, basically because of the sharpness of the BAO peak which around that scale gives $x^2 \xi''(x)/\xi(x) \sim 150$ and ${\Delta_\psi(x) / (x^2\; \xi(x))}\sim 3$.  This means that this correction is much larger than they dynamical correction which is of order $\xi(x)^2$~\footnote{{Since $\xi(x)$ receives contribution also from modes $k\gg 1/x$, the size of the dynamical effects is partially controlled by $\xi(x')$ with $x'\ll x$.}}  
\be
\Delta\xi^{\rm 1-loop}(x)\sim   \xi''(x) \Delta_\psi(x) \sim \xi(x) \gg \xi(x)^2,
\ee
at the BAO scale. This corrections is quite large and thus one should keep many orders in the expansion to get an accurate answer. Thus in this paper we will present formulas in which the exponential is not expanded and thus all the effects in $\Delta_\psi(x)$ are included. In the real case, there are dynamical effects of the long modes and those we will treat perturbatively. 

Another important thing to keep in mind is that, as the toy model above illustrates, the effect of $\Delta_\psi(x)$ is to reduce the contribution of the high $k$ modes to the correlation function, thus broadening the peak. But, as is well known, most of this effect will be substantially reproduced by the so-called reconstruction procedure that tries to undo precisely these motions~\cite{Eisenstein:2006nk}.

This paper is mainly concerned with the corrections to the power spectrum around the non-linear scales, thus in scales which are naively much smaller than the BAO scale. Even so, resumming the terms proportional to $\epsilon_{s<}$ will be very important. In fact the existing calculation of the one~\cite{Carrasco:2012cv} and  two-loop~\cite{Carrasco:2013mua} power spectrum in the EFT showed a striking pattern of oscillations in the residuals which we reproduce later in Figure \ref{2loop}. 
We will show that this residuals are related to $\epsilon_{s<}$ and disappear once that parameter is resumed. 
 
Again to gain some intuition we can investigate the problem in our toy model. 
The one-loop power spectrum is just the Fourier transform of the one-loop correlation function. It is given by:
\be
P^{\rm 1-loop}(k) = P_{13} + P_{22} = -  \langle(\psi(0))^2\rangle k^2 P_L(k) + \int {dk^\prime \over 2 \pi} P_\psi(k^\prime) (k-k^\prime)^2 P_L(k-k^\prime)\ . 
\ee
This can be written in a suggestive way:
 \be
P^{\rm 1-loop}(k) =  \int {dk^\prime \over 2 \pi} P_\psi(k^\prime) [(k-k^\prime)^2 P_L(k-k^\prime)- k^2  P_L(k)]\ . 
\ee
This expression clearly shows the cancelation in the limit $k^\prime \rightarrow 0$.  This is the source of the standard  intuition that modes with  $k^\prime \ll k$ are not relevant for the power spectrum at wavenumber $k$. This would lead to the wrong conclusion that long modes are not relevant for the damping of the BAO peaks at high $k$-values in the power spectrum. 

The cancelation of the effect from modes with $k^\prime \ll k$ comes about through the cancellation of the $P_{13}$ and $P_{22}$ contributions, one that is a convolution and one that is not. So if $P_L$ has some characteristic scale, the cancelation only happens for modes of scale larger than that characteristic scale. In particular, the BAO feature in real space corresponds to oscillations in the power spectrum with a characteristic scale of 100 Mpc. Thus $P_L$ contains a terms that roughly looks like $\cos[k/\Delta k_{\rm oscillations}]$, with $\Delta k_{\rm oscillations}\sim 2\pi/100\,$Mpc$^{-1}$, and it is only for modes $ k^\prime \ll \Delta k_{\rm oscillations}$, rather than $ k^\prime \ll  k$, that the cancellation between $ [(k-k^\prime)^2 P_L(k-k^\prime)- k^2  P_L(k)]$ happens. The motions produced by modes $\Delta k_{\rm oscillations} < k^\prime < k$ still have an effect on the oscillatory part of the power spectrum. 

The standard intuition that the motions produced by modes with $ k^\prime \ll  k$ should not enter can be explained in another way which will also highlight the physical reason why it is not the case for the BAO oscillations. One is tempted to think that to measure the power spectrum at a scale $k$ one only needs a region of size $1/k$. One can then imagine changing coordinates and describing the physics in the local inertial frame. The physics inside that region cannot depend on the motion of the region produced by long modes but only on the second derivatives of the gravitational potential~\cite{Carrasco:2013sva}. This logic is correct but misses the BAO oscillations. If one takes a region of size $1/k$ one has a resolution in $k$ of order $k$ and thus cannot see the oscillations with scale $\Delta k_{\rm oscillations} \ll k$. Thus the argument cannot be used to understand what happens to those oscillations as the local observer is blind to them. This is rather obvious if one thinks in real space, as to be able to measure the BAO feature one needs a region larger than $100$ Mpc. If one is discussing such a large box, the equivalence principle argument only applies for modes that are uniform over this large region. 

The rest of the paper is organized as follows: in section \ref{resum} we present our new resummation technique, in section \ref{PS} we show our power spectrum results and in \ref{REAL} we discuss the density real space correlation function. We conclude in \ref{CONC}.

\section{A simple method to resum the IR effects}\label{resum}

As it was described in~\cite{Porto:2013qua}, the simplest way to describe the EFTofLSS is in terms of a fluid-like continuum of extended particles, characterized by multipoles of their energy distribution, which move under gravity and source gravity according to their extendedness.  The multipoles encompass the effect of the short distance physics that has become non-linear and cannot be described by an analytical treatment. 

As it was described in the former section, for the particular shape of the power spectrum that happens to be in our universe, there are important effects that come from long wavelength modes acting on short, but still mildly linear, modes. The effects from the non-linear physics scale with the parameters $\epsilon_{\delta<}$ or $\epsilon_{s>}$, as powers of $k/\knl$, with $k$ being the wavenumber of a given mode and $\knl$ the wavenumber associated to the non-linear scale.  There are additional IR effects controlled by the parameters $\epsilon_{s<}$, which become of order one at a lower $k$. If these IR effects were to be treated perturbatively, they would introduce a new perturbative expansion parameter, $\epsilon_{s<}$, which is different from $\epsilon_{\delta<}\propto k/\knl$ and of order one, and this would make the perturbative convergence to the true answer much slower. For this reason, it is convenient to resum them, which means to treat the effect of the long distance modes on the mildly non-linear ones non perturbatively. 

This is what we are going to do in the following section. We will derive results first neglecting the extendedness of the particle-like objects that describe the EFTofLSS, and then reintroducing the effect of their extendedness only at the very end. This two-step treatment is possible because we are interested in resumming the IR effects, while the extendedness of the objects is relevant for correctly taking into account the UV physics and so is irrelevant for the IR resummation. While we could in principle resum the  IR effects immediately on the full theory, neglecting the extendedness in the first step allows us to streamline many of the relevant formulas.

Assuming point-like particles, in real space the density and momentum divergence fields are given by:
\bea
1+\delta(\vec r,t) &=& \int d^3 q \ \delta^D(\vec r- \vec q - \vec s(\vec q,t)) \nonumber \ , \\
\pi(\vec r,t) &=& \vec\nabla_r\cdot \int d^3 q \  \dot{\vec s} (\vec q,t) \delta^D(\vec r- \vec q - \vec s(\vec q,t))\ , 
\eea
where for simplicity we defined the $\pi$ field as
\be\label{eq:pi_delta}
\pi(\vec r,t)=-\dot\delta 
\ee
Notice that, by the continuity equation, $\pi$ is related to the divergence of the momentum $\pi^i$:
\be
\rho_b\,\dot\delta+\frac{1}{a}\, \d_i\pi^i=0\ \qquad \Rightarrow\qquad\pi(\vec r,t)=- \frac{1}{a \rho_b} \d_i\pi^i\ ,
\ee
with $\rho_b$ being the background density. The corresponding fields in Fourier space, for $\vec k\neq0$, are: 
\bea
 \delta(\vec k,t) &=& \int d^3 q\; \exp[- i \vec k \cdot  (\vec q+\vec s)]  \nonumber \ , \\
 \pi(\vec k,t) &=& i \int d^3 q \  \vec k \cdot \dot{\vec s} (\vec q,t) \exp\left[- i \vec k \cdot (\vec q + \vec s(\vec q,t))\right] \ .
\eea

To compute correlations easily, let us define
\be
\mu(\vec k,t,\lambda) = \int d^3 q \ \exp\left[- i \vec k \cdot (\vec q + \vec s(\vec q,t) - \lambda\, \dot{\vec s} (\vec q,t) )\right]\ , 
\ee
such that, for $\vec k\neq 0$,
\be
 \delta(\vec k,t) = \left. \mu(\vec k,t,\lambda)\right|_{\lambda=0}\ ,\qquad\qquad
 \pi(\vec k,t) = {d \over d \lambda} \left. \mu(\vec k,t,\lambda)\right|_{\lambda=0}\ . 
\ee

The power spectrum of the density is therefore given by:
\be
\langle   \delta(\vec k_1,t_1)  \delta(\vec k_2,t_2) \rangle =  (2\pi)^3 \delta^{(3)}( \vec k_1+\vec k_2) \int d^3 q \ e^{- i\, \vec k_1 \cdot \vec q}\; \langle\exp\left[ i\, \vec k_1 \cdot \vec{\Delta}(\vec q;t_1,t_2)  \right]\rangle\ ,
\ee
where we have defined 
\be
\vec{\Delta}(\vec q;t_1,t_2) = \vec s(\vec q,t_1)- \vec s(\vec 0,t_2)\ .
\ee 
If we call 
\be
X(\vec k_1,\vec q;t_1,t_2) =  \vec k_1 \cdot \vec{\Delta}(\vec q;t_1,t_2)  \ ,
\ee 
then by the cumulant theorem we have
\be
\langle\exp\left[ i \vec k_1 \cdot \vec{\Delta}(\vec q;t_1,t_2)  \right]\rangle = \exp\left[\sum_{N=1}^{\infty} {i^N \over N! } \langle X(\vec k_1,\vec q;t_1,t_2)^N \rangle_c\right]\ , 
\ee
where $\langle\ldots\rangle_c$ stays for the connected part of the correlation function. 
As a result we find:
\be\label{eq:delta}
\langle   \delta(\vec k_1,t_1)  \delta(\vec k_2,t_2) \rangle =  (2\pi)^3 \delta^{(3)}( \vec k_1+\vec k_2) \int d^3 q \ e^{- i \vec k_1 \cdot \vec q}\; \exp\left[\sum_{N=1}^{\infty} {i^N \over N! } \langle X(\vec k_1,\vec q;t_1,t_2)^N \rangle_c\right]\ . 
\ee
One can easily see that the equal time density power spectrum is IR safe by noticing that the expectation value involves only $\vec{\Delta}(\vec q;t_1,t_2) = \vec s(\vec q,t_1)- \vec s(\vec 0,t_2)$. A mode $k$ that is constant over the separation of the two points, $k\, q \ll 1$, cancels in the difference of displacements, if the displacements are evaluated at the same time, and leads to an additional $k^2$ in the calculation of the  $\vec{\Delta}$ correlation functions. As we discussed in the former section, this means that modes with $k\ll \Delta k_{\rm oscillations}$ will give a vanishingly small contribution to the power spectrum. The situation is different for unequal time matter power spectra, where a mode $k$ that is constant over the separation of the two points, $k\, q \ll 1$, still contributes to $\vec{\Delta}(\vec q;t_1,t_2) \simeq \vec s(\vec 0,t_1)- \vec s(\vec 0,t_2)$. 

We can also compute correlation involving the momentum. In order to to that, we start by computing correlation functions involving $\mu$:
\bea
&&\langle   \delta(\vec k_1,t_1)  \mu(\vec k_2,t_2,\lambda_2) \rangle =  (2\pi)^3 \delta^{(3)}( \vec k_1+\vec k_2)\\
&&\qquad\qquad \int d^3 q \ e^{- i \vec k_1 \cdot \vec q}\; \langle\exp\left[ i \vec k_1 \cdot \vec{\Delta}(\vec q;t_1,t_2)  - i \lambda_2 \vec k_1 \cdot \dot{\vec s} (\vec 0,t_2)]\right]\rangle \nonumber\ , \\[0.4cm]
&&\langle   \mu(\vec k_1,t_1,\lambda_1)  \mu(\vec k_2,t_2,\lambda_2) \rangle =  (2\pi)^3 \delta^{(3)}( \vec k_1+\vec k_2) \nonumber \\ 
&&\qquad\qquad \int d^3 q \ e^{- i \vec k_1 \cdot \vec q}\;\langle\exp\left[ i \vec k_1 \cdot \left(\vec{\Delta}(\vec q;t_1,t_2)  + (  \lambda_1  \dot{\vec s} (\vec q,t_1)-\lambda_2  \dot{\vec s} (\vec 0,t_2))\right)\right]\rangle\ .  \nonumber
\eea
In order to obtain the momentum correlation we need to take derivatives with respect to $\lambda_1$ and $\lambda_2$ and then set  those $\lambda$s to zero. Thus we only need to compute the expectation values to linear order in both $\lambda_1$ and $\lambda_2$ (which includes the cross term $\lambda_1 \lambda_2$). To compute the derivatives with respect to $\lambda$s we can use that the $\lambda$ dependence comes from the replacement
\be
X(\vec k_1,\vec q;t_1,t_2)\rightarrow X(\vec k_1,\vec q;t_1,t_2) +  \vec k_1\cdot (  \lambda_1  \dot{\vec s} (\vec q,t_1)-\lambda_2  \dot{\vec s} (\vec 0,t_2))
\ee
Thus, the derivatives can be computed as
\bea
\left.{d \over d\lambda_1}\ldots\;\right|_{\lambda_1=0}&=&  \vec k_1\cdot \dot{\vec s} (\vec q,t_1) \left.{d \over dX}\ldots\;\right|_{\lambda_1=0}\ , \nonumber \\
\left.{d \over d\lambda_2}\ldots\;\right|_{\lambda_2=0}&=& -  \vec k_1\cdot \dot{\vec s} (\vec 0,t_2)\left. {d \over dX}\ldots\;\right|_{\lambda_2=0}\ . 
\eea
Thus, for the cross correlation term, we get
\bea\label{eq:deltapi}
\langle   \delta(\vec k_1,t_1)  \pi(\vec k_2,t_2) \rangle &=& - (2\pi)^3 \delta^{(3)}( \vec k_1+\vec k_2)  \int d^3 q \ e^{- i \vec k_1 \cdot \vec q} \exp\left[\sum_{N=1}^{\infty} {i^N \over N! } \langle X(\vec k_1,\vec q;t_1,t_2)^N \rangle_c\right]
\nonumber \\
&&\times\;\left[\sum_{N=1}^{\infty} {i^N \over (N-1)! } \langle X(\vec k_1,\vec q;t_1,t_2)^{N-1}   \vec k_1\cdot \dot{\vec s} (\vec 0,t_2) \rangle_c\right]\ . 
\eea
For the momentum auto spectrum we have
\bea\label{eq:momentum}
&&\langle   \pi(\vec k_1,t_1)  \pi(\vec k_2,t_2) \rangle = - (2\pi)^3 \delta^{(3)}( \vec k_1+\vec k_2)  \int d^3 q \ e^{- i \vec k_1 \cdot \vec q} 
\exp\left[\sum_{N=1}^{\infty} {i^N \over N! } \langle X(\vec k_1,\vec q;t_1,t_2)^N \rangle_c\right]
\nonumber \\ &&
\left\{ \left[\sum_{N=1}^{\infty} {i^N \over (N-1)! } \langle X(\vec k_1,\vec q;t_1,t_2)^{N-1}\,   \vec k_1\cdot \dot{\vec s} (\vec q,t_1) \rangle_c\right]  \left[\sum_{N=1}^{\infty} {i^N \over (N-1)! } \langle X(\vec k_1,\vec q;t_1,t_2)^{N-1} \,  \vec k_1\cdot \dot{\vec s} (\vec 0,t_2) \rangle_c\right] \nonumber\right. \\  &&
+\left.  \left[\sum_{N=2}^{\infty} {i^N \over (N-2)! } \langle X(\vec k_1,\vec q;t_1,t_2)^{N-2} \,  \vec k_1\cdot \dot{\vec s} (\vec q,t_1)\,\vec k_1\cdot \dot{\vec s} (\vec 0,t_2) \rangle_c\right]  \right\}\ .
\eea

In this derivation so far we have not expanded in the displacement $\epsilon_{s<}$ as a small parameter. This is where for example the exponential of correlation functions of $X$ that appears in eq.s~(\ref{eq:delta}), (\ref{eq:deltapi}), and (\ref{eq:momentum}) plays a role. Using these expressions as they are for actual computations is however quite inconvenient. It is clear that we do not wish to resum exactly the displacement field, which would include corrections in $\epsilon_{s>}$ and $\epsilon_{\delta<}$. Indeed this would be impossible to do exactly, because the displacement field receives contributions from short distance fluctuations that we control only as an expansion in perturbation theory. Luckily, we can content ourselves with resumming just the long wavelength part of the displacement, which is dominated by IR modes and therefore can be treated linearly. This allows us to perform the calculation using a trick, as we are now going to explain. 

\subsection{Matter-Matter and Matter-Momentum Power Spectra}

Let us first describe the procedure for the matter power spectrum of~(\ref{eq:delta}). Of all the terms in the exponent, only the first one does not contain any power of $\epsilon_{\delta<}$ or $\epsilon_{s>}$, and is therefore the one associated purely to the contribution from the displacements, controlled by $\epsilon_{s<}$. In the Eulerian perturbation theory, we treat on equal footing the parameter that control the IR-displacement, $\epsilon_{s<}$, and the ones that control the  density perturbations $\epsilon_{\delta<}$ and the UV displacements $\epsilon_{s>}$. We simply call them perturbations.  If we were to Taylor expand a Lagrangian expression in powers of the displacement field as well, and treat powers of $\epsilon_{s<},\ \epsilon_{s>}$ and $\epsilon_{\delta<}$ on the same footing, then we would find that the expression we obtain to order $N$ would agree with the corresponding Eulerian expression to order~$N$. 

This suggests the following procedure. Let us call $K(\vec k,\vec q;t_1,t_2)$ the expression in parenthesis in~(\ref{eq:delta}):
\be\label{eq:kdef}
K(\vec k,\vec q;t_1,t_2)= \exp\left[\sum_{N=1}^{\infty} {i^N \over N! } \langle X(\vec k_1,\vec q;t_1,t_2)^N \rangle_c\right]\ .
\ee 
We are interested in evaluating this expression at all orders in the linear long-wavelength displacement fields $\epsilon_{s<}$, and to order $N$ in powers of $\epsilon_{\delta<}$ or $\epsilon_{s>}$. Since we are  going to resum neither in $\epsilon_{\delta<}$ nor in $\epsilon_{s>}$, we can treat these two parameters as the same, and lets us denote them simply as $\epsilon_{\delta<}$. Let us denote an expression evaluated {\it up to} order $N$  in $\epsilon_{\delta<}$, and all orders in $\epsilon_{s<}$, by the following
\be
\left. K(\vec k,\vec q;t_1,t_2) \right|_N \ .
\ee 
Instead, let us denote the same expression evaluated {\it up to} order $N$ by expanding both in $\epsilon_{\delta<}$ and in  $\epsilon_{s<}$, and counting them on equal footing, as
\be
\left. \left. K(\vec k,\vec q;t_1,t_2) \right|\right|_N \ .
\ee 
Let us now define as $K_0(\vec k_1,\vec q;t_1,t_2)$ the following quantity
\be\label{eq:ko}
 K_0(\vec k,\vec q;t_1,t_2) =\exp\left[ -\frac{1}{2} \langle X_0(\vec k,\vec q;t_1,t_2)^2\rangle\right]
\ee
where $X_0$ is the expression $X$ evaluated with the linear solutions. Since we are interested in resumming only the linear displacements, $X_0$ contains all the relevant information we wish to resum out of the exponential in~(\ref{eq:kdef}). We will give details in App.~\ref{app:smoothing-terms} on how to compute this term. Once we have $K_0$, we can use it to do the following manipulations, valid up to order $N$ in $\epsilon_{\delta<}$. Since at all orders in $\epsilon_{s<}$ and zeroth order in $\epsilon_{\delta<}$, $K$ and $K_0$ are equal, we can  multiply and divide by $K_0$, and Taylor expand $K/K_0$ in powers of $\epsilon_{\delta<}$ and $\epsilon_{s<}$. In formulas, we have
\bea\nonumber\label{eq:kapprox}
&&\left. K(\vec k,\vec q;t_1,t_2) \right|_N \simeq K_0(\vec k,\vec q;t_1,t_2)\cdot\left.\left. {K(\vec k,\vec q;t_1,t_2)\over K_0(\vec k,\vec q;t_1,t_2)}\right|\right|_N \nonumber\\ 
&&\qquad\qquad\qquad=K_0(\vec k,\vec q;t_1,t_2) \sum_{j=0}^N \left.\left.K_0^{-1}(\vec k,\vec q;t_1,t_2)\right|\right|_{N-j} K(\vec k,\vec q;t_1,t_2)_{j}  \nonumber\\  
&&\qquad\qquad\qquad=  \sum_{j=0}^N \left( K_0(\vec k,\vec q;t_1,t_2) \cdot\left.\left.K_0^{-1}(\vec k,\vec q;t_1,t_2)\right|\right|_{N-j} \right)\cdot K(\vec k,\vec q;t_1,t_2)_{j} \ .
\eea
Here the subscript ${}_j$ means that we take the order $j$ in $\epsilon_{\delta<}$ {\it and} $\epsilon_{s<}$ of a given expression, no to be confused with~${}||_j$, which  means instead that we take all terms of the same expression {\it up to} order~$j$ in~$\epsilon_{\delta<}$ {\it and} $\epsilon_{s<}$. We remind that expanding at a given overall order in $\epsilon_{\delta<}$ {\it and} $\epsilon_{s<}$ is nothing but the usual expansion in powers of the power spectrum, where we do not distinguish between factors of  $\epsilon_{\delta<}$ and of $\epsilon_{s<}$.  

The final result of the above expression has the following useful property.  By construction, the two expressions agree up to order $N$ in $\epsilon_{\delta<}$, differing only for terms of  order higher than $N$ in $\epsilon_{\delta<}$, that we do not compute anyway. This is so because,  if we take $\epsilon_{s<}=0$, up to order $N$ in $\epsilon_{\delta<}$ the two expressions are identical by construction, as all terms up to order $N$ from $K_0$ have been designed to cancel identically. 

The approximate formula (\ref{eq:kapprox}) agrees with $\left. K(\vec k,\vec q;t_1,t_2) \right|_N$ to all order in $\epsilon_s$ in the limit in which the long displacements are treated as free. This is another useful property of the above expression. We realize that this is the case by analyzing the terms that we are not capturing with the approximate treatment of $K$. The terms that we did neglect originate from the terms that are associated with the connected part of the $N$-point functions that are exponentiated in $K$. Since those $N$-point functions are connected, they need to be made so by insertion of interactions proportional to $\epsilon_{\delta<}$. These insertions come into two classes: either we insert powers of $\epsilon_{\delta<}^{\rm high}$, that is powers of $\epsilon_{\delta<}$, with $\epsilon_{\delta<}$ being evaluated at those high-$k$ modes that we are interested to compute, or we insert powers of $\epsilon_{\delta<}^{\rm low}$, corresponding to the size of $\epsilon_{\delta<}$ in the infrared which dominate the contribution to $\epsilon_{s<}$.  Therefore, these connected terms scale in different ways in powers of $\epsilon_{s<}$, $\epsilon_{\delta<}^{\rm high}$ and $\epsilon_{\delta<}^{\rm low}$. There is no point in resumming the terms just containing powers of $\epsilon_{\delta<}^{\rm high}$, without additional powers of $\epsilon_{\delta<}^{\rm low}$ or $\epsilon_{s<}$, because there are terms of equal order that we do not compute. However, there are terms that have mixed powers of $\epsilon_{s<}$, $\epsilon_{\delta<}^{\rm high}$ and $\epsilon_{\delta<}^{\rm low}$, and, thanks to $\epsilon_{s<}$, they can be bigger than the terms with the same number of power spectra, but containing only  $\epsilon_{\delta<}^{\rm high}$. The most important of these are the ones that account for the non-linear evolution of the displacement fields. For example, $\langle X^4\rangle_c$ contains terms that go as $\epsilon_{s<}^2\epsilon_{\delta<}^{\rm low}$. Smaller terms, that also appear in  $\langle X^4\rangle_c$,  are  for example proportional to $\epsilon_{s<}\epsilon_{\delta<}^{\rm low}\epsilon_{\delta<}^{\rm high}$. It makes sense to resum these terms, with the following procedure. Luckily, at each order in this three-parameter expansion, only a subset of the correlation functions $\langle X^n\rangle_c$ contribute, as the minimum number of power spectra is given by the integer part of $(n+1)/2$, and to make the correlation function connected, one considers that  fluctuations evaluated at low $k$, where they count as $\epsilon_{s<}$, are connected to themselves or to others evaluated at high $k$, where they count as $\epsilon_{\delta<}^{\rm high}$, by internal lines that count as $\epsilon_{\delta<}^{\rm low}$. Only high-$k$ fluctuations are connected by powers of $\epsilon_{\delta<}^{\rm high}$. In practice, to perform the resummation of the terms that are not maximal in powers of $\epsilon_{\delta<}$, one sustitutes $K_0$ with
\bea
&& K_0(\vec k,\vec q;t_1,t_2) =\exp\left[ -\frac{1}{2} \langle X_0(\vec k,\vec q;t_1,t_2)^2\rangle\right]\quad\to\quad  \\ \nn
 &&\  K_0(\vec k,\vec q;t_1,t_2) =\exp\left[ -\frac{1}{2} \langle X_0(\vec k,\vec q;t_1,t_2)^2\rangle_c-\frac{i}{3!}\langle X_0(\vec k,\vec q;t_1,t_2)^3 \rangle+\frac{1}{4!}\langle X_0(\vec k,\vec q;t_1,t_2)^4 \rangle_c+\ldots\right] ,
 \eea
where each of the terms $\langle X_0(\vec k,\vec q;t_1,t_2)^{2,3,4,\ldots}\rangle$ are evaluated to the relevant order one wishes to resum {(therefore  $X_0$ is now constructed with the long wavelength fields and no longer evaluated just on the linear solution)}. When performing this procedure, relevant counterterms from the Lagrangian EFT should be included. It should be stressed that, even with only the resummation of the linear displacements, the series is convergent, because at this point higher order terms can be formed only by adding terms that contain powers of $\epsilon_{\delta<}\ll1$. The resummation of higher order terms is therefore useful only in order to make the perturbative expansion more rapidly convergent, so that the residual mistake is dominated by terms that scale with powers of only $\epsilon_{\delta<}^{\rm high}$. In this paper we will limit ourself to resumming the displacement fields only in the limit in which they are free. Indeed the corrections coming from the non-linearities in the displacements, that can be resummed as we just described, are actually very small. The leading terms that we do not resum are of order $\epsilon_{\delta<}^{\rm high}(\epsilon_{\delta<}^{\rm low})^{n-1}\epsilon_{s<}^n$, with $n\geq 2$, if we do a two-loop calculation, or with $n\geq 1$, if we do a one-loop calculation. These start in size as respectively three-loop or two-loop terms in the counting of the Eulerian EFT, but as just one-loop term in the counting of the Lagrangian EFT counting. Fortunately, $\epsilon_{\delta<}^{\rm low}$ can be estimated to be numerically of order $(\epsilon_{\delta<}^{\rm high})^2$, so, quantitatively, these correspond to approximately a two-loop term in the Lagrangian EFT~\footnote{More quantitatively, the difference between the true displacements and the displacements in the Zeldovich approximation, which is very similar to the ones we resum here, have been shown to be of order percent~\cite{Tassev:2013rta}.}. In this paper we compute only one quantity at two-loops, the equal-time matter power spectrum. Since this is IR-safe, the resummation only affects the oscillations, which are very small, or order $2\%$, to start with. The residual corrections, as we will see, are very small, even though higher order computations can be in principle performed, as we have just explained.

We finally make an additional comment. Our resummation applies only to the long wavelengths modes. Of course, there is some displacement coming also from the short wavelength modes, which gives a qualitatively similar contribution, but that is smaller numerically. We do not resum those terms, which means that this effect originating from the displacements will be reconstructed only order by order in perturbation theory. This leads to an partial resummation. Therefore one can think that  after our resummation, the parameter controlling the effect of the displacement that appear in perturbation theory is no more $\epsilon_{s<}$, which is of order one in $\Lambda$CDM, but rather a new $\epsilon_{s<}$, dubbed $\tilde \epsilon_{s<}$, which is much smaller than one, and for which perturbation theory actually converges, in the usual sense of asymptotic series.

 The effectiveness of formula (\ref{eq:kapprox}) can be explicitly verified in the simplest cases. We can write  the order $N=1$ and $N=2$ expressions from above.
For $N=1$ we have
 \be
 \left. K(\vec k,\vec q;t_1,t_2) \right|_1\simeq1+ K_0^{-1}(\vec k,\vec q;t_1,t_2)\; K(\vec k,\vec q;t_1,t_2)_1\ ,
 \ee
 where, if we Taylor expand the exponential in $K_0$, we have
 \be
  \left. K(\vec k,\vec q;t_1,t_2) \right|_1\simeq 1+K(\vec k,\vec q;t_1,t_2)_1-\frac{1}{2} \langle X_0(\vec k,\vec q;t_1,t_2)^2\rangle K(\vec k,\vec q;t_1,t_2)_1+\dots\ .
 \ee
 The first two terms are exactly   $\left.\left. K(\vec k,\vec q;t_1,t_2) \right|\right|_1$ while the remaining ones are higher order in $\epsilon_{\delta<}$ or~$\epsilon_{s<}$. Therefore, if we do not Taylor expand the exponential, the approximate expression differs from the correct one just by higher terms in $\epsilon_{\delta<}^{\rm high}$ or $\epsilon_{\delta<}^{\rm low}$, as we wished to verify. Similarly, for $N=2$ we have
 \bea
  &&\left. K(\vec k,\vec q;t_1,t_2) \right|_2\\ \nonumber
  &&\qquad\simeq1+ K_0(\vec k,\vec q;t_1,t_2) \left[\left(K_0^{-1}(\vec k,\vec q;t_1,t_2)_0+K_0^{-1}(\vec k,\vec q;t_1,t_2)_1\right)K(\vec k,\vec q;t_1,t_2)_1+K(\vec k,\vec q;t_1,t_2)_2\right] \\ \nonumber
  &&\qquad=1+  e^{-\frac{1}{2} \langle X_0(\vec k,\vec q;t_1,t_2)^2\rangle}\left[\left(1+\frac{1}{2} \langle X_0(\vec k,\vec q;t_1,t_2)^2\rangle\right)K(\vec k,\vec q;t_1,t_2)_1+K(\vec k,\vec q;t_1,t_2)_2\right]
 \eea
 If we Taylor expand the exponential, we can see that only higher order terms in $\epsilon_{s<}$ or $\epsilon_{\delta<}$ survive:
 \bea
  &&\left. K(\vec k,\vec q;t_1,t_2) \right|_2\\ \nonumber
  &&\qquad\simeq1+  K(\vec k,\vec q;t_1,t_2)_1+K(\vec k,\vec q;t_1,t_2)_2\\ \nonumber
  &&\quad\qquad+{\cal O}\left(\langle X_0(\vec k,\vec q;t_1,t_2)^2\rangle^2 K(\vec k,\vec q;t_1,t_2)_1,\; \langle X_0(\vec k,\vec q;t_1,t_2)^2\rangle K(\vec k,\vec q;t_1,t_2)_2 \right)\\ \nonumber
  &&\qquad\simeq \left.\left. K(\vec k,\vec q;t_1,t_2)\right|\right|_{2}+{\cal O}\left(\langle X_0(\vec k,\vec q;t_1,t_2)^2\rangle^2 K(\vec k,\vec q;t_1,t_2)_1,\; \langle X_0(\vec k,\vec q;t_1,t_2)^2\rangle K(\vec k,\vec q;t_1,t_2)_2 \right)
 \eea
Again, if we do not Taylor expand the exponential, the approximate expression differs from the correct one just by higher terms in $\epsilon_{\delta<}$, or terms involving $\epsilon_{\delta<}^{\rm low}$, as we wished to verify.

Let us plug expression (\ref{eq:kapprox}) in (\ref{eq:delta}):
\bea\label{eq:delta2} \nonumber
&&\left.\langle   \delta(\vec k_1,t_1)  \delta(\vec k_2,t_2) \rangle\right|_{N} \\
&&   \qquad=(2\pi)^3 \delta^{(3)}( \vec k_1+\vec k_2) \int d^3 q \ e^{- i \vec k_1 \cdot \vec q} \;\left.\exp\left[\sum_{N=1}^{\infty} {i^N \over N! } \langle X(\vec k_1,\vec q;t_1,t_2)^N \rangle_c\right]\right|_{N} \\ \nonumber 
&&\qquad\simeq (2\pi)^3 \delta^{(3)}( \vec k_1+\vec k_2)\\ \nonumber
&&\quad\qquad \int d^3 q \;e^{- i \vec k_1 \cdot \vec q}\;   \sum_{j=0}^N\left[ \left( K_0(\vec k_1,\vec q;t_1,t_2) \cdot\left.\left.K_0^{-1}(\vec k_1,\vec q;t_1,t_2)\right|\right|_{N-j} \right)\cdot K(\vec k_1,\vec q;t_1,t_2)_{j}\right]\ . 
\eea
If we now define 
\be
F_{||_{N-j}}(\vec k,\vec q;t_1,t_2)=K_0(\vec k,\vec q;t_1,t_2) \cdot\left.\left.K_0^{-1}(\vec k,\vec q;t_1,t_2)\right|\right|_{N-j} \ ,
\ee
the expression for the matter power spectrum simplifies to
\bea\label{eq:delta3} 
&&\left.\langle   \delta(\vec k_1,t_1)  \delta(\vec k_2,t_2) \rangle\right|_{N} \\  \nonumber
&&\qquad\qquad =(2\pi)^3 \delta^{(3)}( \vec k_1+\vec k_2) \int d^3 q \  e^{- i \vec k_1 \cdot \vec q}\;  \sum_{j=0}^N\left[ F_{||_{N-j}}(\vec k_1,\vec q,t_1,t_2)\cdot K(\vec k_1,\vec q;t_1,t_2)_{j}\right]\ .  \nonumber 
\eea

This expression becomes quite intuitive if we go to real space and compute the matter correlation function
\be\label{eq:delta4} 
\left.\xi_{\delta\delta}(\vec r;t_1,t_2)\right|_N=\sum_{j=0}^N\int \frac{d^3 k}{(2\pi)^3}\int d^3 q\; e^{-i \vec k\cdot (\vec q-\vec r)}\;F_{\left.\right||_{N-j}}(\vec k,\vec q;t_1,t_2) \; K(\vec q,\vec k;t_1,t_2)_j
\ee
It is  useful to manipulate the above expression by multiplying by 1 written as
\be
1=\int \frac{d^3 k'}{(2\pi)^3}\; (2\pi)^3\delta^{(3)}(\vec k'-\vec k)=\int\frac{d^3 k'}{(2\pi)^3} \int d^3 q' \; e^{i\, \vec q'\cdot(\vec k'-\vec k)} \ .
\ee
We can at this point replace some $\vec k$ in (\ref{eq:delta4}) with $\vec k'$, to obtain
\bea
&&\left.\xi_{\delta\delta}(\vec r;t_1,t_2)\right|_N=\\ \nonumber
&&\qquad \sum_{j=0}^N\int \frac{d^3 k}{(2\pi)^3}\int \frac{d^3 k'}{(2\pi)^3}\int d^3 q\int d^3 q'\; e^{-i \vec k'\cdot (\vec q'-\vec r)}\;F_{||_{N-j}}(\vec q,\vec k';t_1,t_2)\; \; e^{-i \vec k\cdot(\vec q-\vec q')}K(\vec q,\vec k;t_1,t_2)_j\ . \nonumber 
\eea
Notice at this point that the integrals would factorize if one of the arguments of $F_{||_{N-j}}(\vec q,\vec k';t_1,t_2)$ was $q'$ instead of $q$. However, we can realize that it is a consistent approximation to replace $q$ with $q'$ in that argument for the following reason. Substituting $q$ with $q'$ amounts to making a mistake proportional to the gradients of the displacements. Since, by construction, all the terms that we add are higher order than $N$ if we were to count powers of $\epsilon_{s<}$ and $\epsilon_{\delta<}$ on equal footing, this corresponds to doing a mistake of higher order in $\epsilon_{\delta<}$, more precisely $\epsilon_{\delta<}^{\rm low}$, beyond the order $N$ at which we work. We therefore can perform the approximate replacement
 \be
 F_{||_{N-j}}(\vec q,\vec k';t_1,t_2) \quad\to\quad F_{||_{N-j}}(\vec q',\vec k';t_1,t_2)\ ,
 \ee
to obtain
\bea\label{eq:delta5} 
&&\left.\xi_{\delta\delta}(\vec r;t_1,t_2)\right|_N\\ \nonumber
&&\qquad=\sum_{j=0}^N\int \frac{d^3 k'}{(2\pi)^3}\int d^3 q'\; e^{-i \vec k'\cdot (\vec q'-\vec r)}\;F_{||_{N-j}}(\vec q',\vec k',t_1,t_2)\int d^3 q \int \frac{d^3 k}{(2\pi)^3}\; e^{-i \vec k\cdot(\vec q-\vec q')}K(\vec q,\vec k;t_1,t_2)_j\ . \nonumber 
\eea
We can now notice that the last term on the former expression is nothing but the $j$-th term of the correlation function if we were to expand both in $\epsilon_{\delta<}$ and in $\epsilon_{s<}$. It is therefore the $j$-th order term we would obtain by doing the calculation in the Eulerian approach. In formulas, the $j$-th order Eulerian correlation function is given by
\be
\xi_{\delta\delta,\,j}(\vec q';t_1,t_2)=\int d^3 q \int \frac{d^3 k}{(2\pi)^3}\; e^{-i \vec k\cdot(\vec q-\vec q')}K(\vec q,\vec k;t_1,t_2)_j
\ee 
By defining a {\it probability} of ending up at physical distance $\vec r$  starting from Lagrangian dinstance $\vec q$ as
\be
P_{||_{N-j}}(\vec r|\vec q;t_1,t_2)=\int\frac{d^3 k}{(2\pi)^3}\; e^{-i \vec k\cdot (\vec q-\vec r)}\;F_{||_{N-j}}(\vec q,\vec k;t_1,t_2)\ ,
\ee
the formula for the correlation function up to order $N$ in $\epsilon_{\delta<}$, and resumed in the displacements (i.e. all orders in $\epsilon_{s<}$), takes the following very simple form
\bea\label{eq:delta6} 
&&\left.\xi_{\delta\delta}(\vec r;t_1,t_2)\right|_N\\ \nonumber 
&&\qquad\qquad=\sum_{j=0}^N \int d^3 q \; P_{||_{N-j}}(\vec r|\vec q;t_1,t_2) \;  \xi_j(\vec q)   =\sum_{j=0}^N \int d q\, q^2 \; P_{{\rm int }||_{N-j}}( r| q;t_1,t_2) \;  \xi_{\delta\delta,\,j}( q,t_1,t_2)\ , \nonumber 
\eea
where in the second passage, using the fact that $\xi(\vec r; t_1,t_2)$ depends only on the modulus $r$ of the distance, we have performed the angular integration over the angles between $\vec r$ and $\vec q$, and we have defined
\be
P_{{\rm int}||_{N-j}}(r|q;t_1,t_2)=2\pi \int_{-1}^{1} d\mu\; P_{||_{N-j}}(\vec r|\vec q;t_1,t_2)\;
\ee
where $\mu=\hat q\cdot\hat r$. Formula (\ref{eq:delta6}) above is very simple: the correlation function up to order $N$ in $\epsilon_{\delta<}$ at distance $\vec r$ is given by a weighted sum of the lower order Eulerian correlation functions at distance~$\vec q$, with weight given by the probability that the Lagrangian distance $\vec q$ ends up at the physical distance~$\vec r$. Indeed, it is straightforward to check that the $P_{||_{N-j}}(\vec r|\vec q;t_1,t_2)$ is normalized
\be
\int d^3r\; P_{||_{N-j}}(\vec r|\vec q;t_1,t_2)=1 \ .
\ee
$P_{||_j}(\vec r|\vec q;t_1,t_2)$ has a typical Guassian-like shape, with width of order 10 Mpc, as expected from simple estimates of the long distance displacements in our universe. In Figure~\ref{fig:probplot} we provide a plot of $P_{{\rm int}||_{N-j}}(r|q;t_1,t_2)$ for $N-j=0,1,2$. Notice that $P_{{\rm int}||_{1,2}}$ are not definite positive, simply in order to avoid overcounting for the probability of a given displacement.

\begin{figure}
\begin{center}
\includegraphics[width=10.2cm]{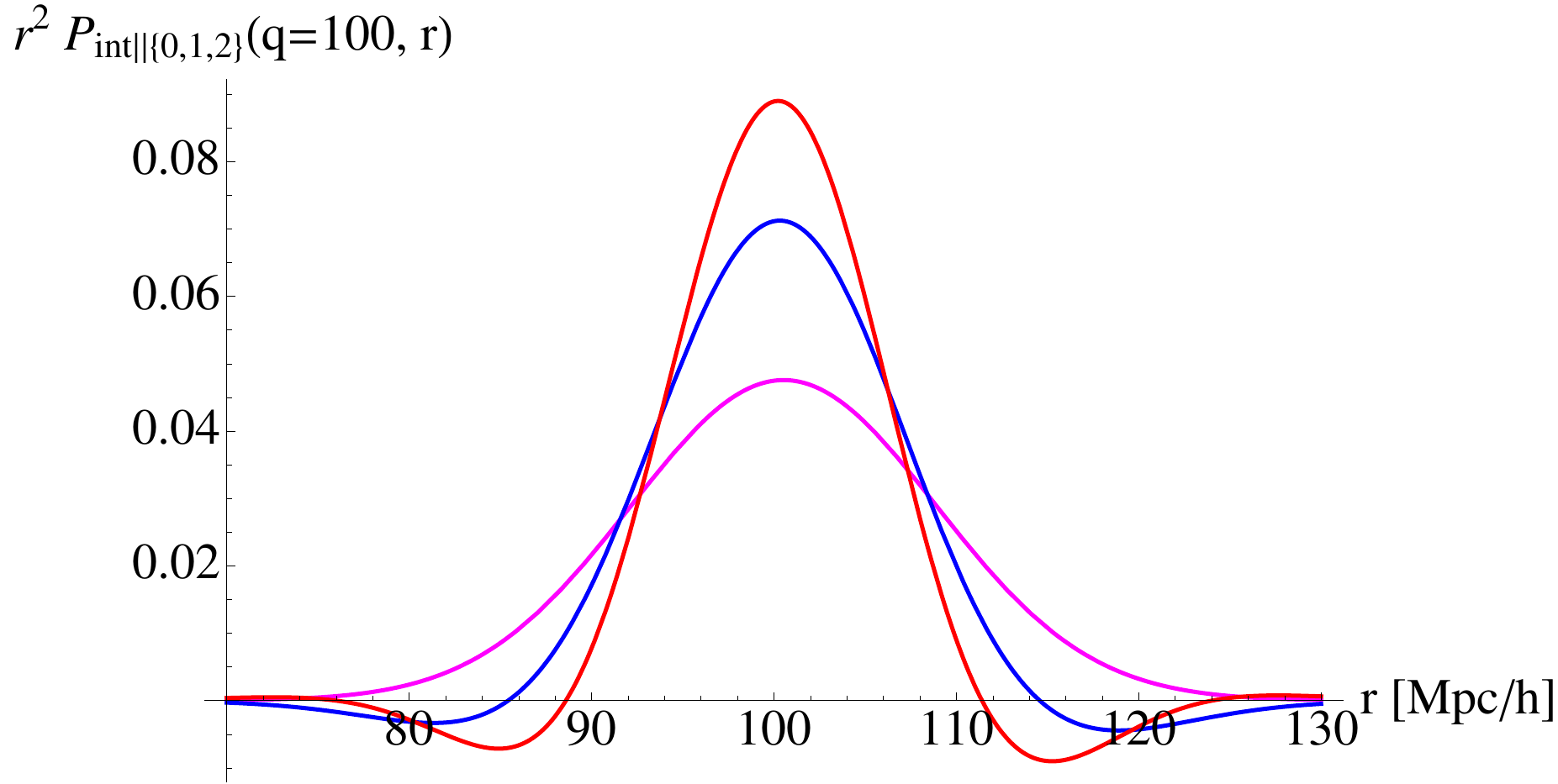}
\caption{\label{fig:probplot} \footnotesize  For $q=100\, {\rm Mpc}/h$, we plot $P_{{\rm int}||_{0}}(r|q;t_1,t_2)$ in magenta, $P_{{\rm int}||_{1}}(r|q;t_1,t_2)$ in blue, and $P_{{\rm int}||_{2}}(r|q;t_1,t_2)$ in red.}
\end{center}
\end{figure}

For numerical reasons, it is actually more convenient to compute directly the power spectrum. Fourier transforming (\ref{eq:delta6}), we have that the matter power spectrum $P_{\delta\delta}(k;t_1,t_2)$, IR-resummed, and up top order $N$ in $\epsilon_{\delta<}$, is given by 
\be\label{eq:delta7} 
\left.P_{\delta\delta}(k;t_1,t_2)\right|_N=\sum_{j=0}^N \int \frac{d^3k'}{(2\pi)^3}\; M_{||_{N-j}}( k, k';t_1,t_2)\; P_{\delta\delta,\,j}(k';t_1,t_2)\ .
\ee
This formula is also extremely simple. The problem of resumming the displacement fields has been reduced to computing the matrixes $ M_{||_{N-j}}( k, k';t_1,t_2)$, that are related to the Fourier transform of the probability of ending up at distance $\vec r$ from distance $\vec q$: 
\be
M_{||_{N-j}}( k, k';t_1,t_2)=\frac{1}{4\pi}\int d^3 r\; d^3 q \; P_{{\rm int}||_{0}}(r|q;t_1,t_2) \; e^{i \vec k \cdot\vec r} \;e^{-i \vec k' \cdot\vec q}\ ,
\ee
and then multiply them by the Eulerian results. In a sense, we have reduced the Lagrangian calculation to become a simple upgrade of the Eulerian one. For equal time matter correlators, in the limit in which $M_{||_{N-j}}( k, k';t_1,t_1)$ is a $\delta$-function in $k$-space, we obtain the usual Eulerian results. This means that the effects of the IR-resummation on the equal-time matter power spectrum can be seen in Fourier space as a mixing of Fourier modes, a sort of convolution in momentum space. As we will see in the next section, this will have very good consequences for reproducing correctly the Baryon Acoustic Oscillations (BAO).

It is straightforward to derive that the same formula holds for the cross correlation of matter and momentum
\be\label{eq:cross2} 
\left.P_{\delta\,\pi}(k;t_1,t_2)\right|_N=\sum_{j=0}^N \int \frac{d^3k'}{(2\pi)^3}\; M_{||_{N-j}}( k, k';t_1,t_2)\; P_{\delta\,\pi,\,j}(k';t_1,t_2)\ ,
\ee
and for the momentum power spectrum
\be\label{eq:momentum2} 
\left.P_{\pi\,\pi}(k;t_1,t_2)\right|_N=\sum_{j=0}^N \int \frac{d^3k'}{(2\pi)^3}\; M_{||_{N-j}}( k, k';t_1,t_2)\; P_{\pi\,\pi,\,j}(k';t_1,t_2)\ .
\ee

So far in this derivation we have neglected all the finite size effects that appear in LEFT and that are associated to the fact that the minimum distances that we hope to describe perturbatively have size of order of the non-linear scale. However, as it is evident by the construction, all these terms contribute in a way that is negligible at large distances, decaying as some powers of $k$ as $k\to 0$. This means that these terms do not play any role in the IR resummation of the effects in $\epsilon_{s<}$, and the effect of the displacement applies to these terms exactly in the same way as it applies to the SPT Eulerian terms. This means that in the EFT, the expressions (\ref{eq:delta7}), (\ref{eq:cross2}) and (\ref{eq:momentum2}) should be meant as simply including in the $P_{\delta\delta,\,j}$, $P_{\pi\pi,j}$ and $P_{\delta\, \pi,\, j}$ terms also the terms that include the counterterms of the Eulerian SPT.

The importance of not expanding in $\epsilon_{s<}$ for $P_{\delta\delta}$ or $P_{\delta\,\pi}$ relies first of all in being able to correctly describe the BAO.  In fact, it is well known (see for example~\cite{Carrasco:2013sva} for a proof using General Relativity arguments), that $P_{\delta\delta}(k;t_1,t_1)$ or $P_{\delta\,\pi}(k;t_1,t_1)$ are IR-safe, which means that modes $k'\ll \Delta k_{\rm oscillations},\; k'\ll k$ do not contribute. 

This indeed can be checked not only by inspection of the Eulerian expressions, but also by inspection of the terms in (\ref{eq:delta}) and (\ref{eq:deltapi}). As we already pointed out, $X(\vec k,\vec q,t_1,t_1)$ is an IR-safe quantity, and this is all what enters in the matter power spectrum. In the matter-momentum cross correlation, we also have
\be\label{eq:cross_ir-safe}
 \langle X(\vec k_1,\vec q;t_1,t_2)^{N-1}\;   \vec k_1\cdot \dot{\vec s} (\vec 0,t_2) \rangle_c\ .
 \ee
At equal times, this vanishes as well for a spatially constant but time dependent displacement. This leads to an additional suppression that makes the equal-time momentum density cross power IR safe.

The situation is different for the momentum autocorrelation. This is not IR safe because it involves the correlation
\be
\langle X(\vec k_1,\vec q;t_1,t_2)^{N-2} \;  \vec k_1\cdot \dot{\vec s} (\vec q,t_1)\;\vec k_1\cdot \dot{\vec s} (\vec 0,t_2) \rangle_c
\ee
which is non-zero even for a displacement with is spatially constant but time dependent. For example the term with $N=2$ in (\ref{eq:momentum}) is simply proportional to 
\be\label{eq:momentum-ir-problem}
\langle \vec k_1\cdot \dot{\vec s} (\vec q,t_1)\,\vec k_1\cdot \dot{\vec s} (\vec 0,t_2) \rangle\ .
\ee
This is contrary to what happened for the equal-time density power spectrum which only involved the correlations of $ X(\vec k,\vec q,t_1,t_1)$.  This means that some additional work might be required in order to perform the IR-resummation for the momentum power spectrum and for the unequal time momentum-matter cross correlation. We are going to address this in the next subsection.

\subsection{Momentum Power Spectrum}

The reason why the resummation of the IR-effects can be done so simply, is because their main effect is simply to displace the location of the extended objects that we try to describe with the EFT. Displacements have little dynamical effect, in the sense that do not deform the objects, they simply translate them.

The formulae that we developed in the former subsection indeed to do not expand in the displacement of the objects. We can think that those equation are supposed to automatically implementing the following conceptual procedure: go to the local inertial frame of the long wavelength displacements, evaluate the dynamics of the short wavelength fluctuations, go back to the original global frame. By the equivalence principle, we know that for long modes $k'\ll \Delta k_{\rm oscillations}$ and $k'\ll k$, where $k$ is the short mode of interest, this treatment encapsulates all the effect. As we described, for modes $k'$ such that $\Delta k_{\rm oscillations}\lesssim k'\ll k$, the effect of the displacement enters also in affecting the size of the oscillations, but this is again correctly taken into account by computing non-perturbatively the displacements themselves.

The formulas we have derived so far do exactly this procedure for the equal-time matter power spectrum and the matter-momentum cross correlation. In the unequal time case, these quantities are not IR-safe, and so the effect of the long displacements is not just simply limited to affecting the oscillations, but they also change the overall size of the correlation function  at a given $k$. The situation is even more complicated for the momentum power spectrum, which is not IR-safe even when evaluated at equal times.

Let us see this in an explicit but schematic way. If we imagine to perform a boost to the local inertial frame of the displacement fields, we perform a change of coordinates of the form
\be
\vec x\ \to\  \vec{\tilde x}= \vec x+\vec s(t)\ ,
\ee
where we have neglected the space dependence of the displacement field. The calculation for the local, tilded, fields is clearly IR-safe, as there are no IR modes in the local inertial frame. Under this change of coordinates, $\delta$ and $\d_i\pi^i$ transform differently:
\be
\delta(\vec x,t)\ \to\ \tilde \delta (\vec {\tilde x},t)=\delta(\vec x(\vec{\tilde x}))\ , \qquad \quad \d_i\pi^i(\vec x,t)\ \to\   \tilde\d_i\tilde\pi(\vec{\tilde x})= \d_i\pi^i(\vec x(\vec{\tilde x}),t)+ \d_i\delta(\vec x(\vec {\tilde x}),t) \cdot \dot s(t)^i\ .
\ee
Both the local $\delta$ and $\d_i\pi^i$ need to be evaluated at the translated coordinates, but $\d_i\pi^i$ is also shifted by a term that is proportional to the gradient of the short wavelength $\delta$, and the velocity of the displacement field. This is nothing but a consequence of the fact that $\delta$ and $\pi^i$ transform differently under change of coordinates. When computing correlation functions, the terms in $\dot s$ cannot be neglected, as they are of order $\epsilon_{s<}$. These terms do not appear in the unequal time matter power spectrum, so eq.~(\ref{eq:delta7}) is correct as is. However, these terms appear both in the momentum power spectrum and in the unequal time matter-momentum cross correlation. In particular, for the matter-momentum cross correlation, we schematically have
\be\label{eq:pi_ir_velocity_cross}
\langle\delta(\vec k_1,t_1) \;\d_i\pi^i(\vec k_2,t_2)\rangle\quad\supset \quad\langle\tilde\delta(\vec{k}_1,t_1)\; k_2^i\, \tilde\delta(\vec k_2,t_2) \, \dot s^i(t_2) \rangle\ .
\ee
This is nothing but a schematic version of eq.~(\ref{eq:cross_ir-safe}), where we see the same term appearing. 
Similarly, for the momentum power spectrum we have
\be\label{eq:pi_ir_velocity}
\langle\d_i\pi^i(\vec k_1,t_1) \;\d_j\pi^j(\vec k_2,t_2)\rangle\quad\supset\quad \langle k_1^i\, \tilde\delta(\vec k_1,t_1) \, \dot s^i(t_2)\; k_2^j\, \tilde\delta(\vec k_2,t_2) \, \dot s^j(t_2) \rangle\ ,
\ee
which is again a schematic version of~(\ref{eq:momentum-ir-problem}). Notice that these terms contain powers of the displacements. For example, in the last expression, we can contract the two long displacements among themselves, without having to pay any gradient suppressions. This implies that in order to properly take into account the effect of the IR-modes for the momentum power spectrum and for the matter-momentum cross correlation at unequal times, it is not enough to carefully keep track of where the extended objects ended up, which is what eq.~(\ref{eq:momentum2}) does, but we need also to keep track of the shifts in the field induced by the displacement velocity. These terms, indeed, are nothing but the terms of the second line of (\ref{eq:deltapi}) and the second and third line of (\ref{eq:momentum}), where we can see that there are terms in the displacement appearing downstairs from the exponential.

Luckily, it is quite easy to improve eq.~(\ref{eq:momentum2}) to take into account of this effect. The reason is that the new terms in $\dot s$ are  not exponentiated, and so they appear only up to a finite maximum power. So, in order to resum them, it is enough not to expand in them when doing the Eulerian calculation. Indeed, eq.s~(\ref{eq:cross2}) and (\ref{eq:momentum2}) are written as a Matrix multiplication acting on some quantities computed in the Eulerian EFT. If we were to find a way to improve the Eulerian result so that the effect of the velocity of the displacement is not expanded perturbatively, than we would be done. Let us do this. 

If we neglect the corrections from the short distance non-linearities, which are irrelevant for the purpose of resumming the IR effects, the Eulerian EFT equations reduce to the SPT equations. Because of the continuity equation~(\ref{eq:pi_delta}), the divergence of the momentum is related to the time-derivative of the matter over density $\pi=-\dot\delta$, so we can study directly $\dot\delta$. Expanding in fluctuations, we have
\be\label{eq:schematic}
\dot\delta=-\frac{1}{a}\d_i\left((1+\delta) v^i\right)=-\frac{1}{a}\left[\theta+\delta\theta+ v^i \d_i\delta\right]\ ,
\ee
where we defined $\theta=\d_i v^i$. If in the last term of the equation above, $\sim v^i\d_i\delta$, we take $v^i$ to be made of long modes, we identify the same IR-large term that we discussed in~(\ref{eq:pi_ir_velocity_cross}) and (\ref{eq:pi_ir_velocity}). This is how the shift in the change of coordinates for non-scalar quantities reveals itself in the Eulerian equations. In the Eulerian EFT computation, all the fluctuations on the right hand side of~(\ref{eq:schematic}) are counted on equal footing, as proportional to $\epsilon_{\delta<}$, but it is clear that  the last term is instead proportional to $\epsilon_{s<}$. For this reason, we should simply consider the term $\sim v^i\d_i\delta$ as simply linear in $\delta$, with no suppression coming from~$v^i$. Terms involving this vertex are the only one in which the IR-resummation is not implemented by the formulas~(\ref{eq:cross2}) and (\ref{eq:momentum2}). The procedure to upgrade them is therefore simply to add the relevant vertexes until enough powers of $\epsilon_{\delta<}$ are included in the calculation. 
 This suggests the following procedure to upgrade eq.~(\ref{eq:cross2}) and and eq.~(\ref{eq:momentum2}):
\bea\label{eq:deltamomentum3} 
&&\left.P_{\delta\,\pi}(k;t_1,t_2)\right|_N\\ \nonumber
&& \quad=\int \frac{d^3k'}{(2\pi)^3}\; \left[\sum_{j=0}^N M_{||_{N-j}}( k, k';t_1,t_2)\;  P_{\delta\,\pi,\,j}(k';t_1,t_2)+ M_{||_{0}}( k, k';t_1,t_2)\;  \Delta P_{\delta\,\pi,\,N}(k';t_1,t_2)\right]\ ,\\ 
\label{eq:momentum3}
&&\left.P_{\pi\,\pi}(k;t_1,t_2)\right|_N\\ \nonumber
&& \quad=\int \frac{d^3k'}{(2\pi)^3}\; \left[\sum_{j=0}^N M_{||_{N-j}}( k, k';t_1,t_2)\;  P_{\pi\,\pi,\,j}(k';t_1,t_2)+ M_{||_{0}}( k, k';t_1,t_2)\;  \Delta P_{\pi\,\pi,\,N}(k';t_1,t_2)\right]\ .
\eea
where,  $\Delta P_{\delta\,\pi,\,N}(k;t_1,t_2)$ and $\Delta P_{\pi\,\pi,\,N}(k;t_1,t_2)$ correspond to adding all terms of order $N$ in $\epsilon_{\delta<}$ that were not included in the standard Eulerian calculation because we considered $v^i$ as a perturbation of order $(\epsilon_{\delta<})^{1/2}$. To give an explicit example, at one-loop order, in the equal-time momentum power spectrum, since $v^i$ appears in the equations at most linear, we need to add only one term, which is given by
\bea \label{eq:momentum_final}
&&\Delta P_{\pi\,\pi,\,N}(k;t_1,t_1)=\frac{1}{a^2} \left[\langle v(\vec x_1,t_1)^i v(\vec x_2, t_1)^j \rangle_{1} \; \langle\d_i \delta(\vec x_1,t_1)\d_j \delta(\vec x_2, t_1) \rangle_{N}\right]_k\\  \nn
&&\qquad\qquad=\frac{1}{a^2}\int^{\bar\Lambda_{\rm Resum}(k)} \frac{d^3 k'}{(2\pi)^3} \;\; P_{\theta\theta,\, 1}(\vec k',t_1,t_1)\; \frac{\left(\vec k'\cdot (\vec k-\vec k')\right)^2}{k'^4} \;P_{\delta\delta,\,N+1}(|\vec k-\vec k'|,t_1,t_1)\ .
\eea
Similar formulas hold at higher orders and similarly for $\Delta P_{\delta\,\pi,\,N}(k;t_1,t_2)$, and can be easily derived if needed.

Here the cutoff $\bar\Lambda_{\rm Resum}(k)$ is a $k$-dependent cutoff that should be taken to be smaller than the $k$'s of interest, for each $k$. For the results presented in this paper we took $\bar\Lambda_{\rm Resum}(k)= k/6$, and we discuss in Appendix~\ref{app:momentum_cutoff} a way to determine the best choice. This cutoff is meant to enforce the fact that we are resumming only long wavelength displacements. The fact that, because of $\bar\Lambda_{\rm Resum}$, we are not resumming the whole of the infrared modes,  implies that the perturbation theory will still be an expansion in an new $\tilde \epsilon_{s<}\ll \epsilon_{s<}$.  Since, if $\bar\Lambda_{\rm Resum}$ is sufficiently large, $\tilde \epsilon_{s<}\ll1$, the remaining perturbative expansion will be convergent, and with it, the residual dependence on $\bar\Lambda_{\rm Resum}$ will become vanishingly small. In practice, for $\bar\Lambda_{\rm Resum}$ sufficiently large, the expansion parameter that controls the perturbative expansion is just $\epsilon_{\delta<}$, which realizes our initial purpose.

Equations~(\ref{eq:delta7}), (\ref{eq:deltamomentum3}) and (\ref{eq:momentum3}) provide the formulas that evaluate the matter and momenta power spectra and cross correlations to order $N$ in $\epsilon_{\delta<}$ and all order in $\epsilon_{s<}$. In the next section, we are going to write explicitly what these equations are up to two-loops for the matter power spectrum, and up to one-loop for the momentum power spectrum and the matter-momentum cross correlation, and show the results when compared with $N$-body simulations. 

\section{Results in Fourier Space}\label{PS}

\subsection{Matter Power Spectrum}

We now present our results. We will limit ourselves to equal time correlation functions.
We start by presenting the matter power spectrum up to two loops. We follow~\cite{Carrasco:2013mua} for the results in the Eulerian EFT. We use the same cosmological parameters $h=0.7136$, $\Omega_{\rm m}=0.258$, $\Omega_{\rm b}=0.0441$, $n_{\rm s}=0.963$, and $\sigma_8=0.796$.  In the region $k\in[0.1,0.7] \invMpc$, we can fit the linear power spectrum of the real universe, as piecewise scaling:
\bea\label{eq:fit}
P_{11}(k)=(2\pi)^3 \left\{ \begin{array}{ll} 
\frac{1}{\knl^3}\left(\frac{k}{\knl}\right)^{-2.1} & \text{ for} \ {k>\ktr} \ , \\
 \frac{1}{\tknl^3}\left(\frac{k}{\tknl}\right)^{-1.7}   & \text{for} \ {k<\ktr} \ ,
\end{array} \right.
\eea
where $\tknl = (\knl^{0.9} \ktr^{0.4})^{1/1.3}$ and $\ktr$ is the transition scale between the two different power-law behaviors.  The fit parameters are
\beq
\knl = 4.6 \invMpc \qquad \ktr = 0.25 \invMpc \qquad \tknl = 1.8 \invMpc \ .
\eeq
In the Eulerian EFT, the one-loop and two-loop matter power spectra take the form
\beq
\label{eq:peft1loop}
P_{\text{EFT-one-loop}} = P_{11} + P_{\text{1-loop}} -  { 2\, (2\pi)} \co \frac{k^2}{\knl^2} P_{11} \ ,
\eeq
while at two loops we have
\beq
P_{\text{EFT-two-loop}}  = P_{11} + P_{\text{1-loop}} +P_{\text{2-loop}}
 -{ 2\,(2 \pi)}  (\co+\ct) \frac{k^2}{\knl^2} P_{11}
 + (2\pi)  \co P_{\text{1-loop}}^{(c_{\rm s}, p)}
 +  (2\pi)^2 c_{s(1)}^4 \frac{k^4}{\knl^4} P_{11} \ .
\eeq
where $\ct$ is a function of $\co$ determined by imposing that at a given renormalization scale $k_{\rm ren}$, $P_{\text{EFT-one-loop}}(\kren) =P_{\text{EFT-two-loop}}(\kren) $:
\be\label{eq:c2equation}
\ct(\kren) = \frac{  P_\text{2-loop}(k_{\rm ren}) + (2\pi) \co(k_{\rm ren}) P_{\text{1-loop}}^{(c_{\rm s})}(k_{\rm ren}) }
{2(2\pi)(k_{\rm ren}^2/\knl^2)P_{11}(k_{\rm ren})} 
+{ \pi [\co(k_{\rm ren})]^2 \frac{k_{\rm ren}^2}{\knl^2} }\ .
\ee

We can now apply the elements of these formulas to the IR-resummed version~(\ref{eq:delta7}). Very explicitly, we have the following formulas. At linear level in $\epsilon_{\delta<}$, we have
\be\label{eq:deltatree} 
\left.P_{\delta\delta}(k;t)\right|_0= \int \frac{d^3k'}{(2\pi)^3}\; M_{||_{0}}( k, k';t)\; P_{\delta\delta,\,11}(k';t)\ .
\ee
At one-loop, we have
\bea\label{eq:deltaoneloop} 
&&\left.P_{\delta\delta}(k;t)\right|_1 \\ \nonumber
&&\; =\int \frac{d^3k'}{(2\pi)^3}\;\left\{ M_{||_{1}}( k, k';t)\; P_{\delta\delta,\,11}(k';t)+\; M_{||_{0}}( k, k';t)\; \left[P_{\delta\delta,\,\text{1-loop}}(k';t)- { 2\, (2\pi)} \co \frac{k^2}{\knl^2} P_{11}\right]\right\},
\eea
while finally at two-loops we have
\bea\label{eq:deltatwoloop} 
&&\left.P_{\delta\delta}(k;t)\right|_2=\int \frac{d^3k'}{(2\pi)^3}\;\left\{ M_{||_{2}}( k, k';t)\; P_{\delta\delta,\,11}(k';t)\right. \\ \nonumber
&&\; +\; M_{||_{1}}( k, k';t)\; \left[P_{\delta\delta,\,\text{1-loop}}(k';t)- { 2\, (2\pi)} \co \frac{k^2}{\knl^2} P_{11}\right]\\ \nonumber
&&\; \left.+\; M_{||_{0}}( k, k';t)\; \left[P_{\delta\delta,\,\text{2-loop}}(k';t)- { 2\, (2\pi)} \ct \frac{k^2}{\knl^2} P_{11}+ (2\pi)  \co P_{\text{1-loop}}^{(c_{\rm s}, p)}
 +  (2\pi)^2 c_{s(1)}^4 \frac{k^4}{\knl^4} P_{11}\right]\right\}.
\eea

Before looking at the results, let us now describe the procedure though which we determine $\co$ and $\ct$. In the one-loop case, the procedure is exactly as in~\cite{Carrasco:2013sva}: we determine $\co$ by simply fitting the predicted power spectrum to the non-linear power spectrum in the range $k\in[0.15,0.25] \invMpc$, where the non-linear data are provided by the Coyote interpolator~\cite{Heitmann:2008eq,Heitmann:2009cu,Lawrence:2009uk,Heitmann:2013bra}. We find a value for $\co$ equal to
\beq  
\co \simeq 1.63 \times \frac{1}{2\pi} \lp \frac{\knl}{\invMpc} \rp^{2} .  
\eeq
This represents a very small shift with the value it was found in~\cite{Carrasco:2013sva}: $\co \simeq  1.62 \times \frac{1}{2\pi} \lp \frac{\knl}{\invMpc} \rp^{2} $.

Instead, the way we determine $\ct$ in the two loop calculation is different than in~\cite{Carrasco:2013sva}. There, we {\it first} determined $\co$ from the one-loop calculation, and then, by using~(\ref{eq:c2equation}), $\ct$ was predicted. That procedure was used in~\cite{Carrasco:2013sva} to minimize the chances of overfitting. However, the procedure is clearly suboptimal: one should rather determine the parameters of the theory using the maximal range of $k$'s available. We do this here by using (\ref{eq:c2equation}) to express $\ct$ in terms of $\co$, and then we fit for $\co$ directly using the two-loop result. The range in $k$'s along which we fit is determined by checking how long we can make the theory curve be parallel to the numerical curve. For the two-loop result, this amounts approximately to the range  $k\in[0.15,0.45] \invMpc$. By using $\kren=0.2 \invMpc$, this procedure leads to
\beq  
\co \simeq 1.65 \times \frac{1}{2\pi} \lp \frac{\knl}{\invMpc} \rp^{2} ,
\eeq
which corresponds to
\be
\ct(\kren=0.2\invMpc) \simeq -3.3 \times \frac{1}{2\pi} \lp \frac{\knl}{\invMpc} \rp^{2} .
\ee
These results are clearly compatible both with the results of~\cite{Carrasco:2013sva}, which found $\ct\simeq -3.3 \times \frac{1}{2\pi} \lp \frac{\knl}{\invMpc} \rp^{2}$ and with the fit at one-loop. It is also quite satisfactory to look at the dependence of $\ct$ on the renormalization scale. If we fix $\co$ to be equal to the best fit value, as we move the renormalization scale in the range $k\in[0.10,0.35]\invMpc$, the value of $\ct$ changes by just $\sim 10\%$ (see Fig.~\ref{cs2ren}). The dependence on the renormalization scale should be indeed small, comparable to the contribution of higher loop terms at those $k$'s . Unfortunately we cannot extend the range of the renormalization scale neither in the IR, as the data become relatively quite noisy with respect to the size of the terms we wish to estimate, as they also become smaller, nor in the UV, as the one-loop result is supposed not to be reliable anymore beyond $k\sim 0.3 \invMpc$.

\begin{figure}
\begin{center}
\includegraphics[width=11.2cm]{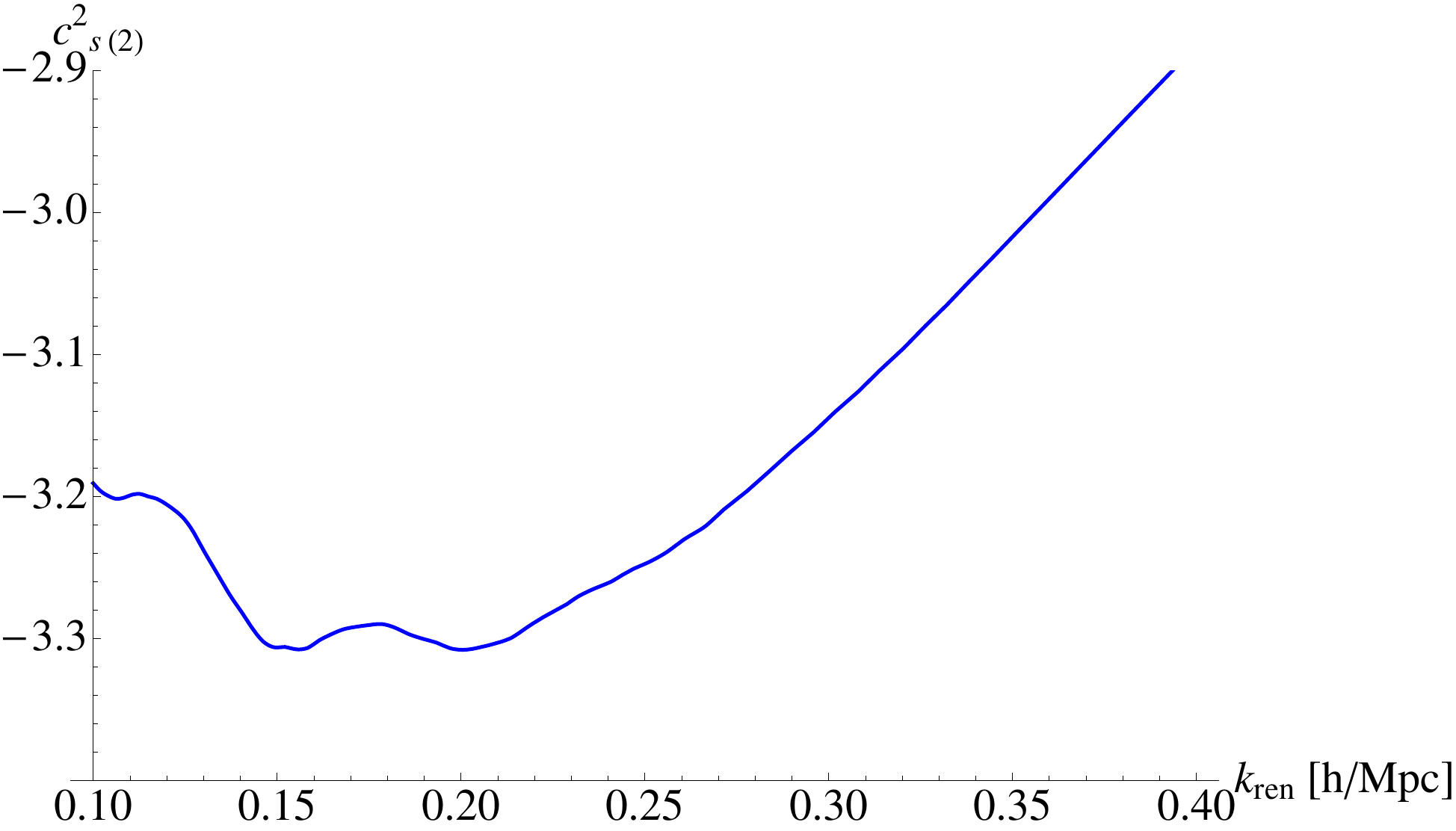}
\caption{\label{cs2ren} \footnotesize  The value of $\ct$ from (\ref{eq:c2equation}) as a function of the renormalization scale $\kren$. We see that as we move $\kren\in[0.10,0.35] \invMpc$, the value of $\ct$ changes by just order $10\%$.}
\end{center}
\end{figure}

We are now ready to look at the numerical results. In the top of Fig.~\ref{2loop} we show the results of the prediction of the IR-resummed EFT at one-loop (in thick red) and two-loops (in thick blue). In thin dashed are represented the results for the Eulerian EFT, that is without IR-resummation, with the same colors respectively. The green band represents the estimated theoretical error from three-loops. The two-loops results have been renormalized at $\kren=0.2\invMpc$. Since the equal-time matter power spectrum is IR-safe, we see that the effect of the IR-resummation is just to affect the oscillations, which are indeed now correctly taken into account. We see that the one-loop result matches to percent level the data up to $k\simeq0.34\invMpc$, while at two-loop matches all the way up to $k\simeq0.6\invMpc$. The spike at $k\simeq 0.05 \invMpc$ is due to the numerical interpolator, against which we compare, not to the EFT. It is however well within the claimed error bars of about percent level. It is also important to notice that the match stops approximately when the three-loop term is estimated to become relevant~\footnote{For the three-loop contribution, we use the estimate~\cite{Carrasco:2013sva} (to which we refer for details): 
\be
P_\text{3-loop}\sim \frac{(2\pi)^2}{2}\left(\frac{k}{\knl}\right)^{2.7} P_{11}(k)\ .
\ee}. This is consistent: since the three-loops is missing from the calculation, it would be unjustified if the two-loops prediction  kept matching the data beyond $k\simeq 0.6\invMpc$.

In the bottom of Fig.~\ref{2loop} we instead compare the results of the IR-resummed EFT with the ones of SPT. In thick magenta, red and blue we plot respectively the IR-resummed linear, one-loop and two-loops predictions of the EFT. With the same colors, but dashed, the same quantities in SPT. As we go to higher orders, SPT does not increase the agreement with the data. This has nothing do with the displacement field, as LPT, which does not expand in the displacements, would have the same reach in the UV. It is simply because both SPT and LPT are ill defined, as they both do not treat properly the UV modes. Furthermore, we notice that SPT has the same residual oscillatory features as the Eulerian EFT. This is due to a lack of resummation of the IR modes with $k_{\rm IR}\gtrsim \Delta k_{\rm oscillation}$. In contrast, the IR-resummed EFTofLSS correctly predicts the size of the oscillations, and, at each order in perturbation theory, it improves the match to the data. Another good property of the EFTofLSS is that at each order in perturbation theory it is possible to estimate in what range of scales the theory should match the data.

{Finally let us comment on the relation between our results and previous ones in the literature. The fact that IR displacements are large and cannot be treated perturbatively in our universe at low redshifts and that they are important in order to correctly reconstruct the BAO oscillations was already pointed out in the literature (see for example~\cite{Eisenstein:2006nk,Crocce:2007dt,Taruya:2012ut,Carlson:2012bu}). Formulas  to resum the contribution of IR modes in perturbation theory have been given in the context of Renormalized Perturbation Theory (RPT)~\cite{Crocce:2005xy} and Regularized Perturbation Theory (RegPT)~\cite{Bernardeau:2011dp,Taruya:2012ut}~\footnote{ The fact that these techniques are named `Renormalized' or `Regularized' should not lead the reader to believe that techniques produce an improvement in the UV with respect to SPT. This would not be correct: they simply differ from SPT in their treatment of the IR modes, not of the UV modes.}. These resummation techniques are very different from ours. In fact these techniques correctly reproduce the BAO peak but they do so in a way that changes the UV behavior of the theory  beyond modifying  the BAO's oscillatory contribution~\cite{Crocce:2007dt,Taruya:2012ut}. In fact, as we argued, the fact that the effect of the IR modes for the equal-time dark-matter power spectrum is limited to the BAO peak (that is to the oscillations of the power spectrum) should reflect itself in the fact that a correctly implemented resummation must have the same UV reach as the theory without resummation. This is not the case for the former resummation techniques~\cite{Crocce:2007dt,Taruya:2012ut}, while it is the case for ours, as it can be seen clearly in Fig.~\ref{2loop}~\footnote{The reader who wants to check this claim more explicitly can compare our eq.~(\ref{eq:delta}) (using (\ref{eq:kapprox})) with eq.~(52) of~\cite{Bernardeau:2011dp}, as suggested to us by the referee. One can clearly see that our formulas are different. In particular, contrary to eq.~(52) of~\cite{Bernardeau:2011dp}, we do not have a Gaussian damping for $\epsilon_{s<}\gtrsim 1$, (in the notation of~\cite{Bernardeau:2011dp},  $k\gtrsim \sigma_{\rm displ.}$).}. This is also why in the bottom of Fig.~\ref{2loop}, we compare against SPT. Former techniques, if correctly implemented, should have the same reach in the UV as SPT, and as LPT, which does not expand in $\epsilon_{s<}$ to start with. As we described, the IR-resummation for non-IR-safe quantities is even more complex.
}

\begin{figure}
\begin{center}
\includegraphics[width=11.2cm]{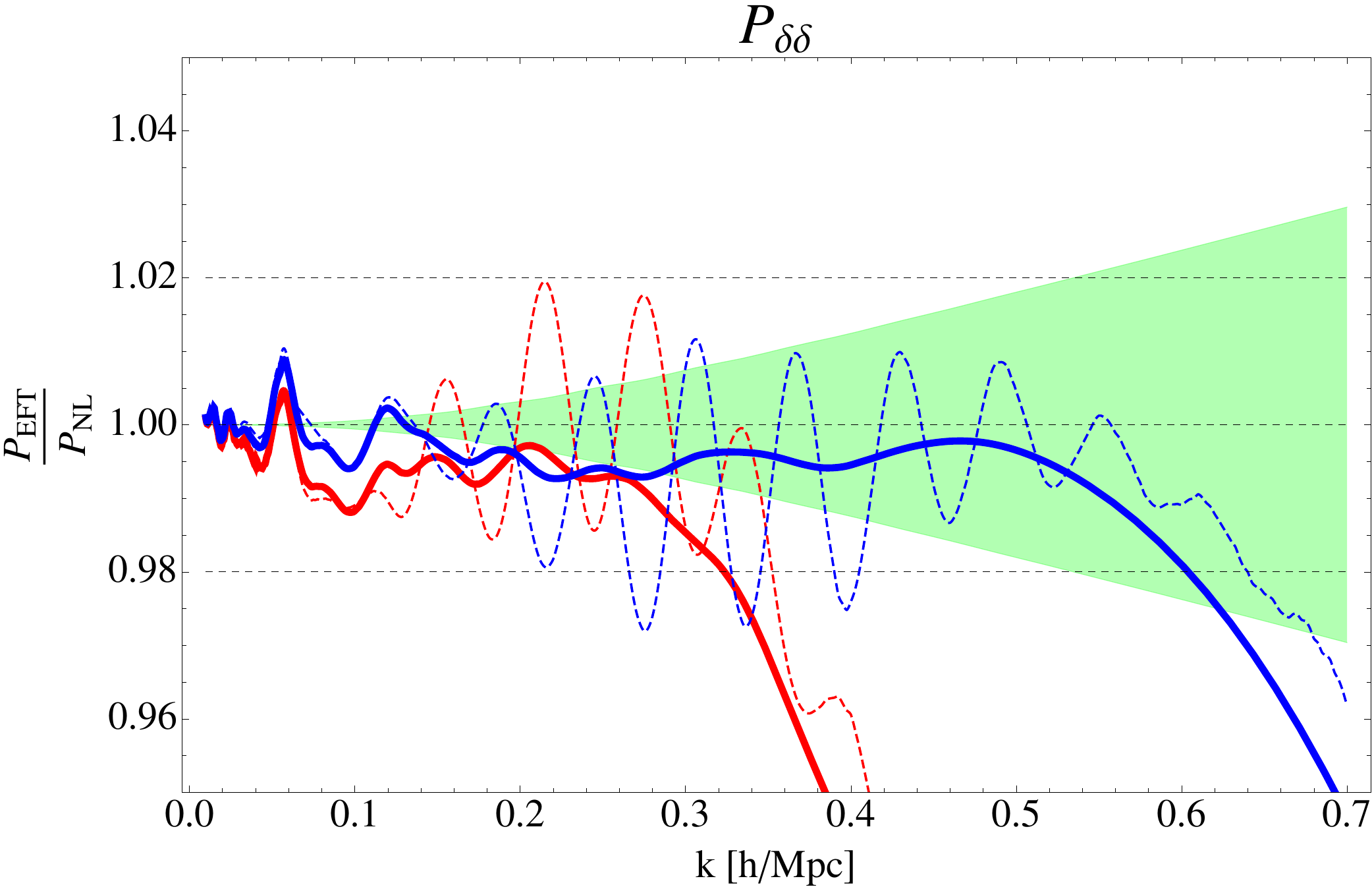}\vspace{0.5cm}
\includegraphics[width=11.2cm]{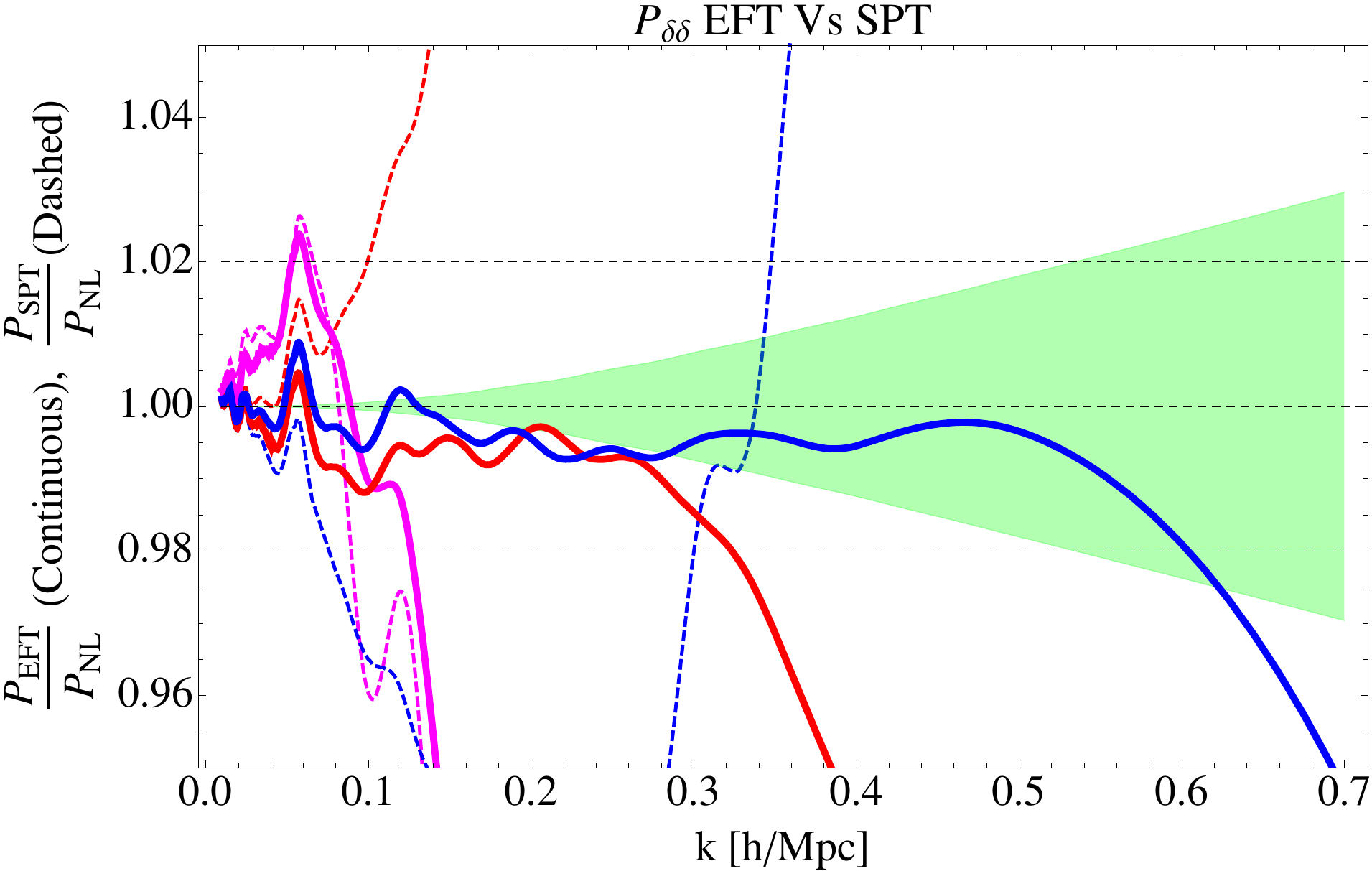}
\caption{\label{2loop} \footnotesize {\it Top: }The prediction of the IR-resummed EFT at one-loop (in thick red) and two-loops (in thick blue). In thin dashed are the predictions from the Eulerian EFT, that is without IR-resummation, with the same colors respectively. The green band represents the estimated theoretical error from three-loops. The two-loops results have been renormalized at $\kren=0.2\invMpc$, and $\co$ has been approximately fit up to $k\simeq 0.5\invMpc$. Since the equal-time matter power spectrum is IR-safe, we see that the effect of the IR-resummation is just to affect the oscillations, which are indeed now correctly taken into account. We see that the one-loop result matches to percent level the data up to $k\simeq0.34\invMpc$, while at two-loop matches all the way up to $k\simeq0.6\invMpc$. The spike at $k\simeq 0.05 \invMpc$ is due to the numerical interpolator, against which we compare, not to the EFT. It is also important to notice that the match stops exactly the three-loop term is estimated to become relevant. 
{\it Bottom:} We compare the predictions of the IR-resummed EFT with the ones of SPT. In thick magenta, red and blue we plot respectively the IR-resummed linear, one-loop and two-loops predictions of the EFT. With the same colors, but dashed, the same quantities in SPT. As we go to higher orders, SPT does not increase the agreement with the data. Furthermore, we notice that SPT has the same residual oscillatory features as the Eulerian EFT.  In contrast, the IR-resummed EFTofLSS correctly predicts the size of the oscillations, and, at each order in perturbation theory, it improves the UV match to the data. Importantly, in the EFTofLSS, order by order in perturbation theory, it is possible to estimate up to where the theory should match the data.}
\end{center}
\end{figure}

\subsection{Correlations Involving Momenta}

We now proceed to illustrate the one-loop results for the equal-time momentum power spectrum and the matter-momentum cross correlation. We use simulations by Okumura {\it et al.}~\cite{Okumura:2011pb}, based on a flat $\Lambda$CDM model with $\Omega_{\rm b}h^2=0.0226$, $\Omega_{\rm m} h^2=0.1367$, $h=0.7$, $n_{\rm s}=0.96$, and $\sigma_8=0.807$. As we discussed, the equal-time momentum matter cross correlation is IR-safe, so that the only effect we expect from resumming the IR modes with respect to the result obtained in the Eulerian EFT is to smooth out the residual oscillations. On the other hand, the momentum power spectrum is not IR-safe, so that we expect that thanks to the IR-resummation not only the oscillations are better described, but also the UV reach is improved. In fact, in~\cite{Carrasco:2013sva} is was shown that, apart for the oscillations, the momentum power spectrum in the Eulerian EFT was affected by strong IR corrections that made the UV reach much smaller than what achieved at the same order for IR-safe quantities. In fact, we now expect that the UV reach of the matter power spectrum, the momentum power spectrum, and the matter-momentum cross correlation, to be quite comparable.

Let us give the relevant expressions. For the matter power spectrum, we use (\ref{eq:peft1loop}), with a different value of $\co$ as the cosmological parameters are now different. For the matter-momentum cross correlation and the momentum power spectrum, we have the following Eulerian EFT formulas~\cite{Carrasco:2013sva}:
\bea
&&
P_{\delta\pi}(a,k) = - \H  \lp
D_1(a) D_1'(a) \, P_{11}(k)
+ 2D_1(a)^3 D_1'(a) \left\{ P_{13}(k) + P_{22}(k) \right\} \right. \\ \nn
&& \left.\qquad\qquad\qquad 
- 2\pi\left(4\,\co\,D_1(a)^3 D_1'(a) +(\co)'\,D_1(a)^4 \right) \frac{k^2}{\knl^2} P_{11}(k) \right) \  ,  \label{eq:pdpi-eds}\\[0.5cm]
&&
P_{\pi\pi}(a,k) = \H  ^2 \lp
D_1'(a)^2 P_{11}(k)
+ D_1(a)^2 D_1'(a)^2 \left\{ 3P_{13}(k) + 4P_{22}(k) \right\} \right. \\ \nn
&&\left. \qquad\qquad\qquad
-  4\pi  \left(3\, \co\,D_1(a)^2 D_1'(a)^2+(\co)'\,D_1(a)^3 D_1'(a)\right) \frac{k^2}{\knl^2} P_{11}(k)\right.\\ \nn
&&\left. \qquad\qquad\qquad+D_1(a)^2 D_1'(a)^2 \Delta P_{\pi\,\pi,\,1}(k;t_0,t_0)\right)  \ , \label{eq:ppipi-eds} \ .
\eea
where the subscript ${}_0$ represents  present time, $'=\frac{d}{da}$, and $D_1$ is the growth factor defined as
\be
\delta(\vec k,a)=\delta(\vec k,a_0)D_1(a)/D_1(a_0)\ .
\ee

We notice that these expressions depend on the time derivative of $\co$. After including $P_{\delta\delta}$, we have three functions with two unknowns, $\co$ and $(\co)'$, and therefore one of these functions has no fitting parameter. We can make things better by using an approximate symmetry. In the limit in which we approximate the universe as Einstein de Sitter, and we consider the power spectrum as a simple power law, there is a scaling symmetry that allows us to determine the time-dependence of $\co$. This is an approximate statement, but, as it is a good approximation to replace Green's function with $D$, we expect it to be sufficiently good in this case. Using this symmetry we have that the time dependence of the $\co$ linear term in the matter power spectrum is~\footnote{For this result, the following reference should be cited as the original derivation~\cite{Simon}.}
\be
P_{\delta\delta}\ \supset\ \co k^2 D_1^2 P_{11} \propto D_1^{2+\frac{4}{3+n}}\ .
\ee
For the range of $k$'s we are interested in at one-loop, we can take $n\simeq -1.7$. In this way, we can predict $P_{\delta\pi}$ and $P_{\pi\pi}$ after using $P_{\delta\delta}$ to determine $\co$.

The formulas for the resummation of the oscillations are analogous to (\ref{eq:deltaoneloop}), with the linear terms being obviously the linear term, and all the rest counting as one-loop terms. For the momentum power spectrum we also add the term in (\ref{eq:momentum_final}), with $N=1$, counted as a one-loop term.

In Figure~\ref{dPidPi} we plot the predictions of the EFT for $P_{\delta\pi}$, $P_{\pi\pi}$ and $P_{\delta\delta}$. In Magenta we have the one-loop SPT, in red the one-loop Eulerian EFT, and in blue we have the IR-resummed EFT with optimized IR-resummation, while in blue dashed we plot the results of the IR-resummed EFT with non-optimized IR-resummation. The band around each line represents the $1$-$\sigma$ cosmic variance of the simulations. Let us start explaining with $P_{\delta\delta}$, where we see that at one-loop, the results are analogous to the ones obtained in the former section. In particular the EFT fits the data up to $k\simeq 0.35\invMpc$. The value of $\co$ we use is
\be
\co \simeq 1.7 \times \frac{1}{2\pi} \lp \frac{\knl}{\invMpc} \rp^{2} .
\ee
We then pass to $P_{\delta\pi}$, where we see that the results are very similar to $P_{\delta\delta}$: the UV reach is about the same, which is a confirmation of the validity of the EFT, as the UV reach should be more or less the same at a given order in perturbation theory for every quantity. Furthermore, the effect of the IR-resummation is simply to better reconstruct the oscillations, as $P_{\delta\pi}$ is an IR-safe quantity. Finally, in $P_{\pi\pi}$ we have two effects. Passing from SPT to the Eulerian EFT, the result agree more with the simulations, but the UV reach is much smaller than for $P_{\delta\delta}$. Passing to the IR-resummed EFT, we see that we achieve two effects. First, the reach in the UV is restored to be approximately the one for matter, and no more no less, as it should be. Second, the oscillations are now correctly computed. 

Finally, let us comment on the difference between the blue lines and the dashed blue lines both in the momentum power spectrum and in the matter-momentum cross power spectrum. As we explain in detail in Appendix~\ref{app:smoothing-terms}, the results with the dashed line are obtained by resumming the displacement fields obtained by summing at linear level the contribution of all the modes up to a cutoff $\Lambda_{\rm resum}=0.1\invMpc$. This is the same resummation we used in the former section for the two-loop matter power spectrum. As we explained earlier, the resummation does not resum the effect of the displacement fields exactly, and leaves us some residual effect from the displacement to be taken into account perturbatively, that is order by order in perturbation theory. Since for the momentum we perform the calculation only up to one-loop, the residual oscillations are smaller than at tree-level, but still clearly still visible. This tells us that the residual $\tilde\epsilon_{s<}$ is smaller than one, but not small enough to make its effect completely negligible at one-loop. As we saw, its effect becomes negligible at two-loops.  It is possible to ameliorate the resummation by noticing the following trick. Since the oscillations are dominated by the displacement fields around the BAO peak, at around $r\sim 120/h$ Mpc, it is a better ansatz for  the resummation to start from a $X_0(q)$ in~(\ref{eq:ko}) that agrees with the one obtained with using the correct displacements at $r\sim 120/h$ Mpc. We therefore rescale the  $X_0(q)$ obtained with modes up to  $\Lambda_{\rm resum}=0.1\invMpc$ to match the correct displacement at $r\sim 120/h$ Mpc. This procedure, dubbed optimized resummation in order to distinguish it from the non-optimized one, amounts to making $\tilde\epsilon_{s<}$ even smaller. We see indeed that the oscillations are almost invisible already at one-loop. We stress that this trick is done just to make the convergence in $\tilde \epsilon_{s<}$ faster, and it does not entails the introduction of new parameters. In fact, the results are quite independent of the particular procedure we decide to implement as an optimized procedure, that is different choices can give the same results. In particular, the optimized and non-optimized resummation will agree already at some low order in perturbation theory.

In summary, the results are quite satisfactory, even though a more accurate check is clearly limited by the large cosmic variance of the simulations~\footnote{A careful reader might wonder why the cosmic variance in the momentum power spectrum is much larger than in the matter power spectrum. This is because indeed the actual simulations are IR-sensitive. The actual result for the momentum power spectrum at short distance does depend strongly on the actual realization that one happens to have for the long modes of the box. This implies that the cosmic variance is strongly affected by the cosmic variance of the long modes, even for short modes. It should be possible to improve the estimator for the momentum power spectrum from measurements, using indeed this insight that we obtained from our analytical calculations.}.

\begin{figure}
\begin{center}
\includegraphics[width=9cm]{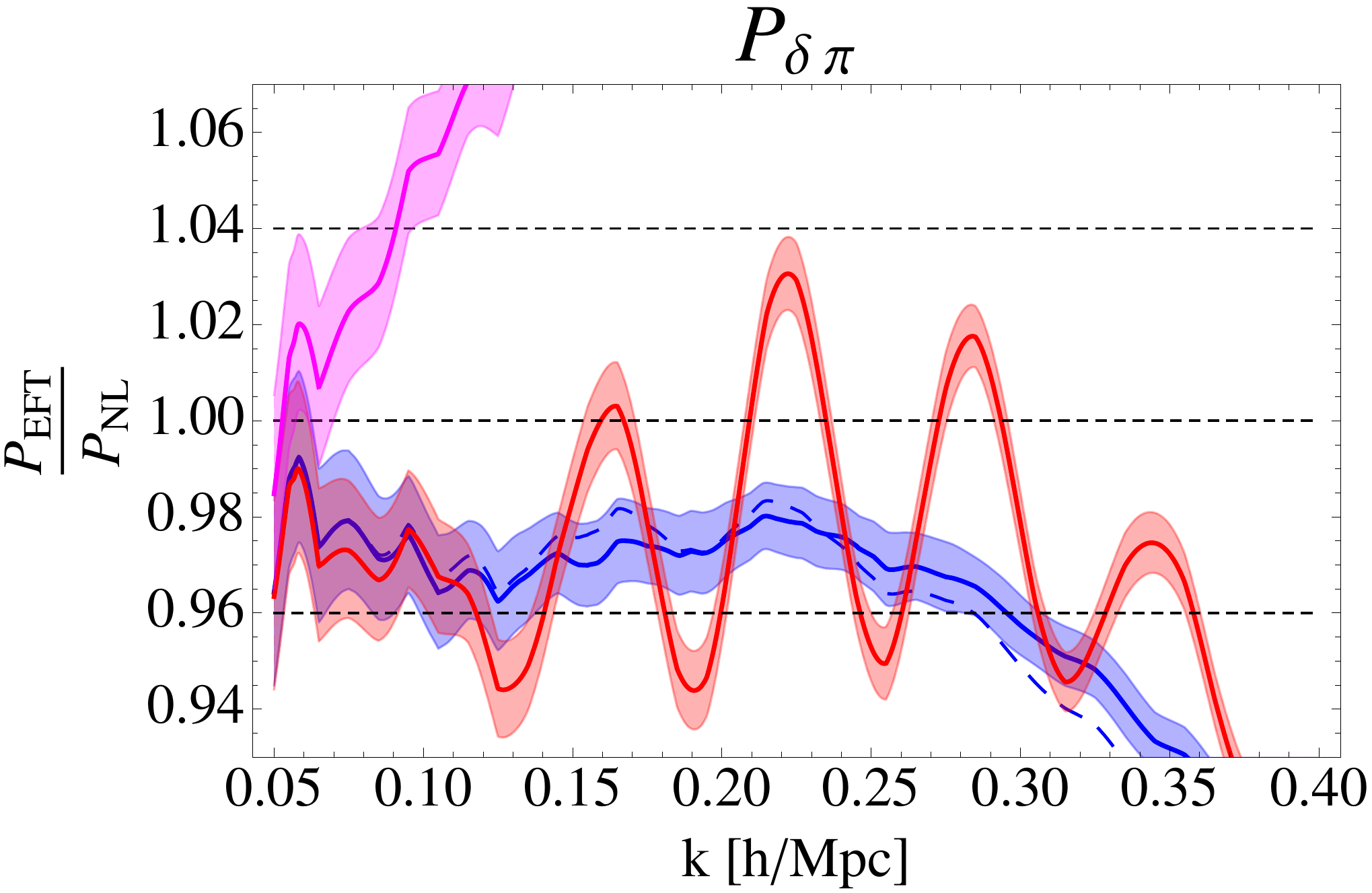}
\includegraphics[width=9cm]{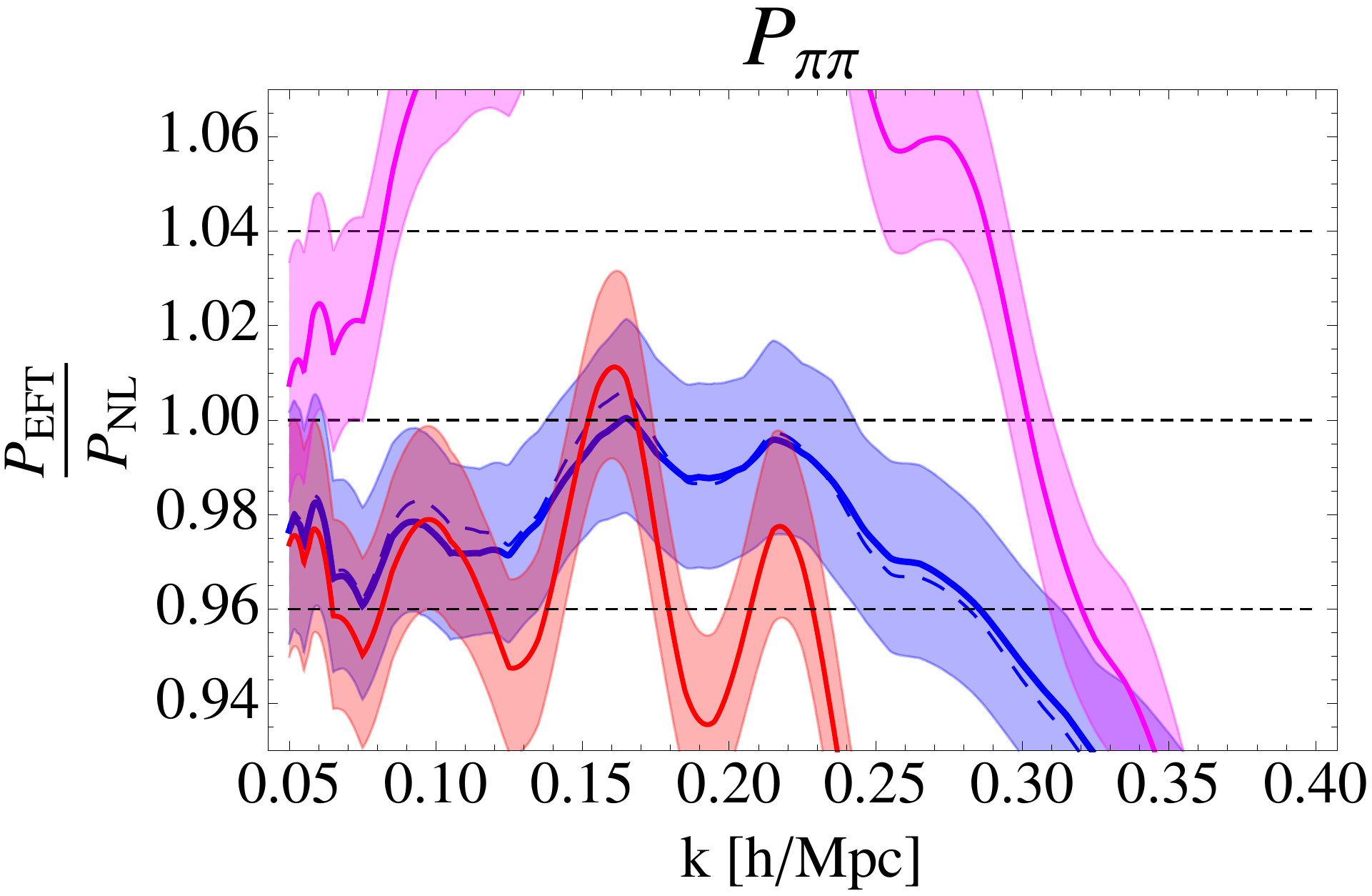}
\includegraphics[width=7cm]{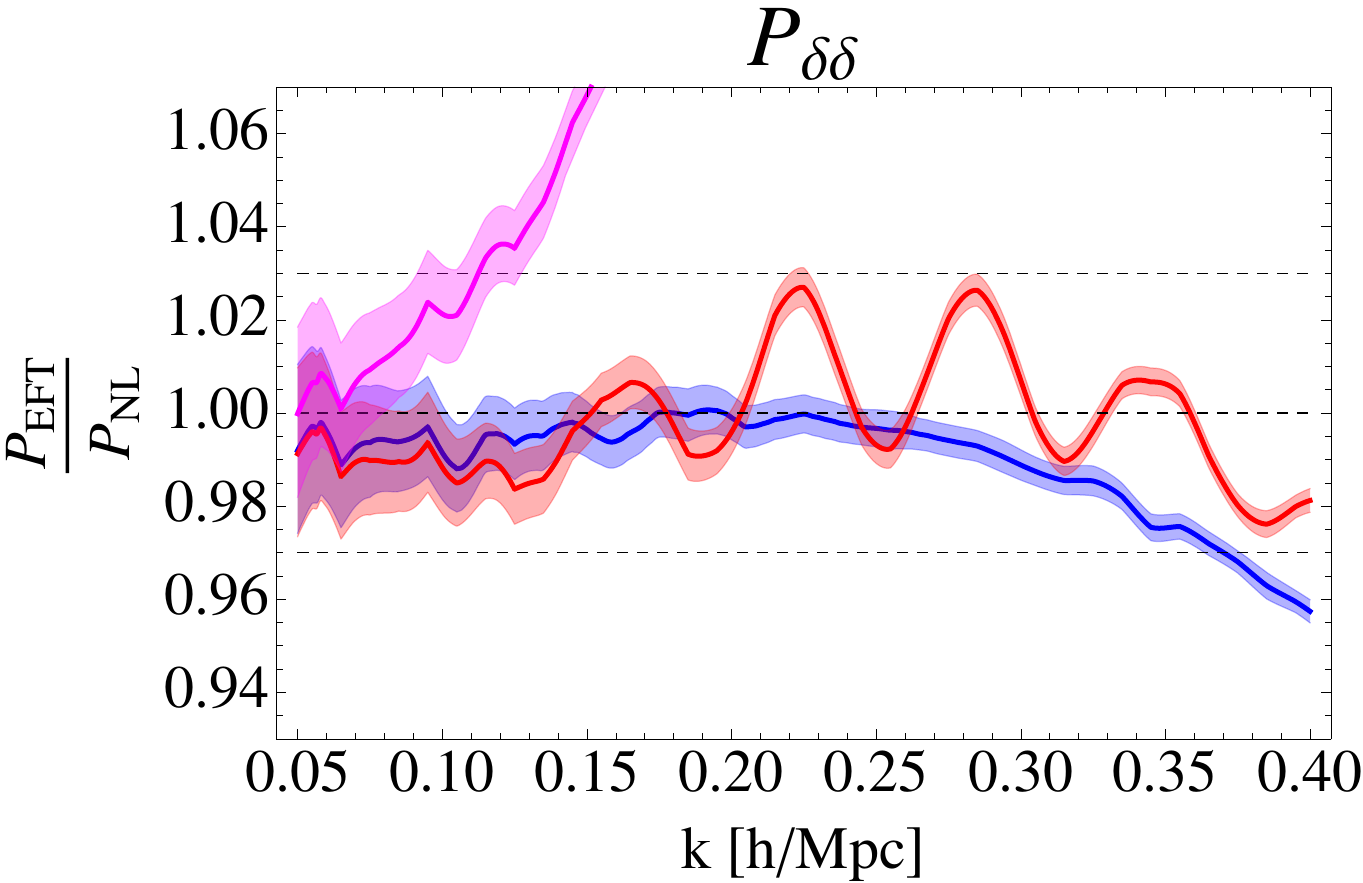}
\caption{\label{dPidPi} \footnotesize  From top, anticlockwise, the predictions of the EFT for $P_{\delta\pi}$, $P_{\pi\pi}$ and $P_{\delta\delta}$. In Magenta we have the one-loop SPT, in red the one-loop Eulerian EFT, and in blue we have the IR-resummed one-loop EFT with optimized IR-resummation, while in blue dashed we plot the results of the IR-resummed one-loop EFT with non-optimized IR-resummation. The band around each line represents the $1$-$\sigma$ cosmic variance of the simulations. For $P_{\delta\delta}$, the results are analogous to the ones obtained in the former section. In particular the EFT fits the data up to $k\simeq 0.35\invMpc$. For $P_{\delta\pi}$, the results are very similar to $P_{\delta\delta}$: the UV reach is about the same, which is a confirmation of the validity of the EFT, as the UV reach should be more or less the same at a given order in perturbation theory for every quantity. Furthermore, the effect of the IR-resummation is simply to better reconstruct the oscillations, as $P_{\delta\pi}$ is an IR-safe quantity. Finally, in $P_{\pi\pi}$ we have two effects. Passing from SPT to the Eulerian EFT, the result agree more with the simulations, but the UV reach is much smaller than for $P_{\delta\delta}$. Passing to the IR-resummed EFT, we see that we achieve two effects. First, the reach in the UV is restored to be approximately the one for matter, and no more no less, as it should be. Second, the oscillations are now correctly computed, especially with the optimized procedure. The results are quite satisfactory, even though a more accurate check is limited by the large cosmic variance of the simulations.}
\end{center}
\end{figure}

\section{Results in Real Space}\label{REAL}

Our Fourier space results can be used to calculate the real space correlation function. This quantity is of great interest because the BAO peak in the correlation function is used as a standard ruler to map the expansion history of the Universe. 

In this paper we will not present comparisons between our analytical results for the correlation function and numerical simulations which we leave for future work. We will present internal consistency checks that we can do based on our results. 

As we have discussed in previous sections, the leading effect changing the shape of the BAO peak are the coherent motions induced by long wavelength modes of wavelengths comparable to the BAO scale which we are resumming in our technique. There are also dynamical effects due to non-linearities. The modeling of these non-linearites improves as we do calculations at higher loops. 

In our resummation technique we keep terms that in the SPT counting would be higher order because they involve high powers of the linear displacement. When we do this resummation we are thus assuming something about long wavelength displacements.  As we have discussed, the successive loop orders correct any mistake we have made in that assumption. Any potential mistake starts at order $P_{11}^{2+N}$ in the $N$-loop calculation. Thus including successive loop orders accomplishes two objectives, it increases the precision with which we include non-linear effects and it decreases the sensitivity on the details of how the resummation was done. 

The EFT consistently tracks the effects of the small scale dynamics that is outside the range of validity of the perturbation theory. It does so by incorporating those effects using a set of free parameters, just $c^2_{s(1)}$ at the order we are working in this paper. The values of those parameters will of course affect the detailed shape of the BAO peak. 

In this section we will compare our results for the real space correlation function around the BAO peak as we increase the loop order. We will investigate its sensitivity to the EFT parameters and to the assumptions we make in the resummation. 

\paragraph{The BAO peak at zero, one and two loops:} Although we will not present a comparison between our results and direct measurements of the correlation function from numerical simulations, in this section we will investigate how the correlation function depends on the loop order we compute to. We will take our highest order computation, the IR-resummed two-loop result, as our best estimate. In Figure~\ref{corrvsloop2} we show the linear theory results as well as the zero-, one- and two-loop results. By 0-loop we mean the linear theory results plus the IR-resummation. 
We also show the relative difference between the successive orders and our best result. While the linear theory result differs from the two-loop one by about 30\%, the 1 and two-loop results are already within a percent of each other. 

Our zero-loop result differs from our best answer by around 10\%. This might be surprising at first. The zero-loop result is basically our version of the Zeldovich approximation. The Zeldovich approximation is known to agree with numerical simulations significantly better than this. The discrepancy can be traced to way we are doing the IR-resummation, in particular the fact that we are  cutting-off the power spectrum at a scale $\Lambda_{\rm resum}=0.1\invMpc$. We will discuss this issue in more detail later.

\begin{figure}
\begin{center}
\includegraphics[width=0.96\textwidth]{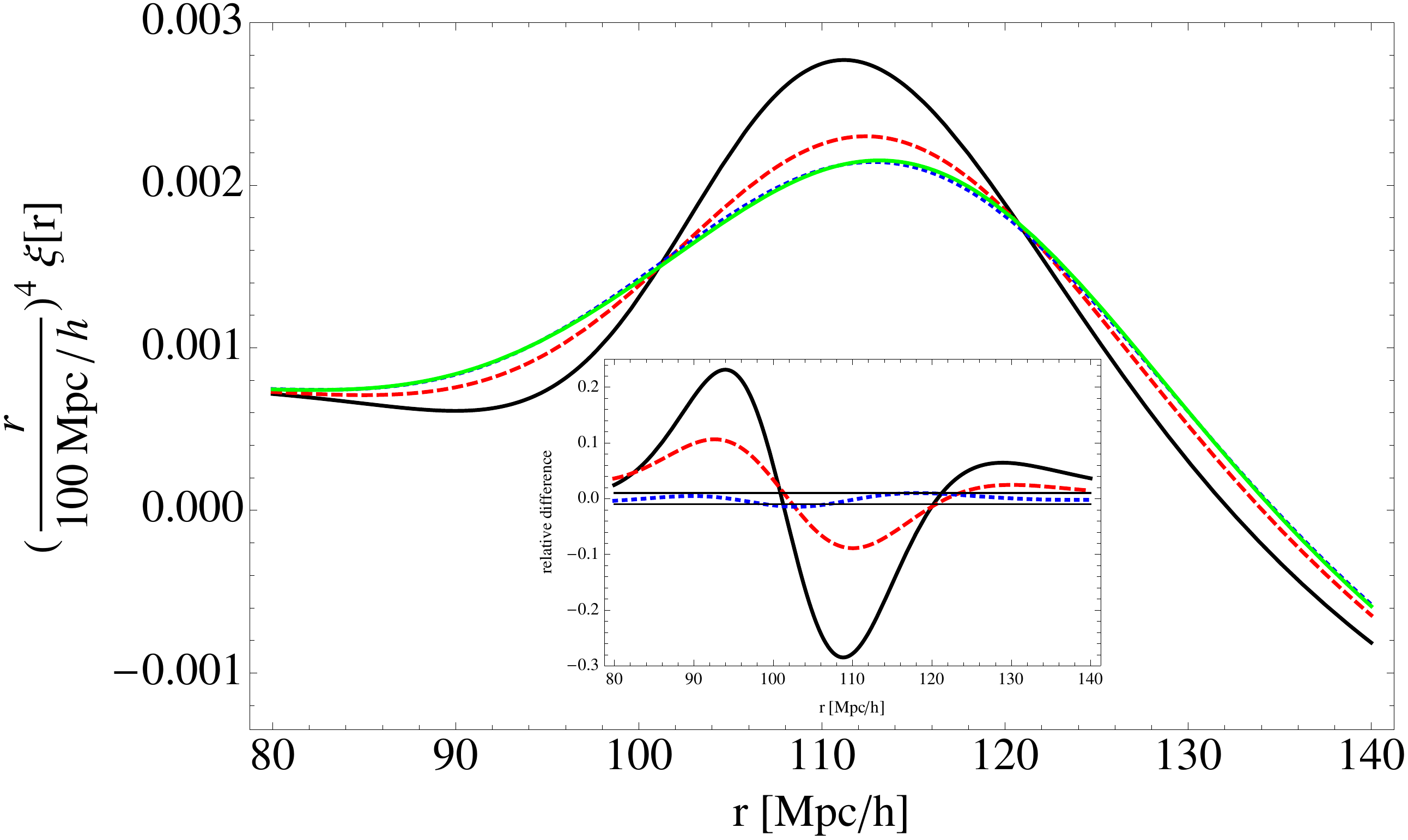}
\caption{\label{corrvsloop2} \footnotesize Correlation function in linear theory (solid black), zero loops (dashed red), one loop (dotted blue) and two loops (solid green). The IR-resummation  was done with $\Lambda_{\rm resum}=0.1\invMpc$. The inset shows the relative difference between different orders and the two-loop result. Because the correlation function goes through zero we define relative difference as $(\xi(r)-\xi_{\text{two-loop}}(r))/\xi_{\text{two-loop}}(r_{BAO}=110 \invMpc)$.} 
\end{center}
\end{figure}

\paragraph{Dependence on $c_s$:} Although the EFT has free parameters we find that at the BAO scale, the shape of the correlation function is very insensitive to the chosen values. We illustrate this in Figure~\ref{cs-dep}. We consider our results at one and two loops with different choices of $c_{s(1)}^2$. In each case we compare a value of $c_{s(1)}^2$ that fits the power spectrum data well with one which does not. In particular we choose the second value such that at the maximum scale where the best fit is still flat as a function of scale, the ``incorrect" $c_{s(1)}^2$ leads to a power spectrum that differs by 3\%, which is well outside of the error bar of the simulation. We see that the resulting differences are extremely small, less than 0.5\% at one loop and less than 0.01 \% at two loops. Of course we do not claim that our results are this accurate but we do conclude that as the EFT parameters  are chosen in such a way as to match the Fourier space statistics, the remaining uncertainties in those parameters do not affect the BAO scale at a relevant level. Equivalently there is no remaining freedom to improve the fit at the BAO scale by changing the EFT parameters without significantly worsening the fit of the power spectrum.

\begin{figure}
\begin{center}
\includegraphics[width=12.2cm]{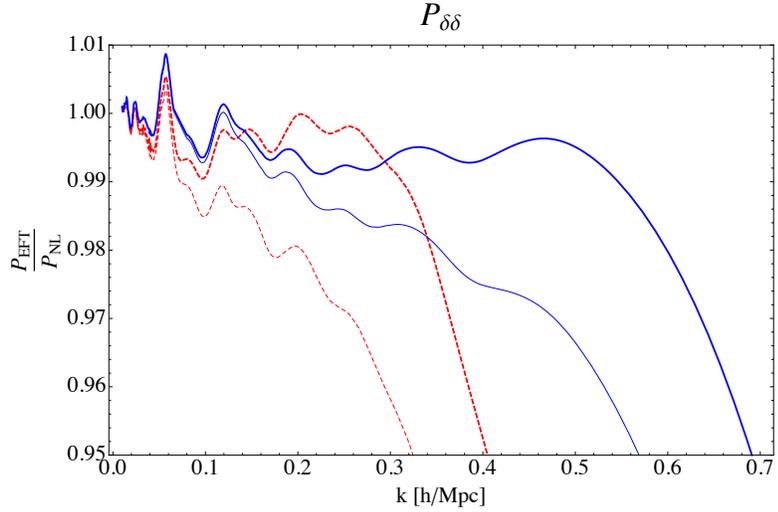}
\includegraphics[width=12.2cm]{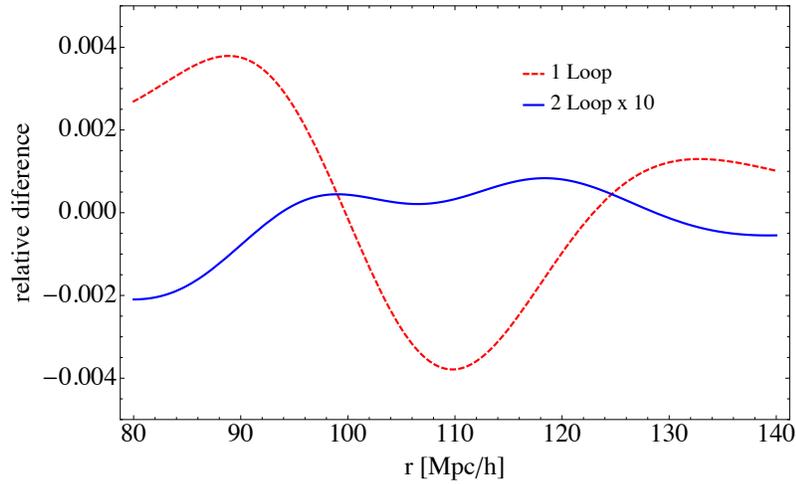}
\caption{\footnotesize  The top panel shows our best results for the one and two-loop power spectra together with results for different values of $c^2_{s(1)}$ chosen such that the resulting power spectrum differs by 3\% with our best value at the maximum scale where the original fit was reliable. The bottom panel shows the resulting relative difference in the correlation function at the BAO scale. The two-loop relative difference was scaled by a factor of 10 to make it more visible. }
\label{cs-dep}
\end{center}
\end{figure}

\paragraph{Dependence on the IR-resummation:} Finally we want to investigate the dependence of the of our results on the details of the IR resummation. As we explain in 
Appendix \ref{app:smoothing-terms}}, we compute $K_0$
\be
K_0(\vec k,\vec q;t) = \exp\left[-\frac{1}{2}A_{ij,1}(\vec q;t) k^i k^j\right] \ ,
\ee
where, by rotational invariance, we must have
\be
A_{ij,\, 1}(\vec q;t)=X(q;t)_1\, \delta_{ij}+ Y(q;t)_1\,\hat q_i \hat q_j\ ,
\ee
with
\bea\label{eq:X}
&& X(q;t)_1=\frac{1}{2\pi^2}\int_0^{+\infty} dk\;  \exp\left[- \frac{k^2}{\Lambda_{\rm Resum}^2}\right]\;P_{\delta\delta,11}(k;t) \left[\frac{2}{3}-2\,\frac{j_1(k q)}{k q}\right]\ , \\
&& Y(q;t)_1=\frac{1}{2\pi^2}\int_0^{+\infty} dk\;\exp\left[- \frac{k^2}{\Lambda_{\rm Resum}^2}\right]\; P_{\delta\delta,11}(k;t) \left[- 2\, j_0(k q)+6\,\frac{j_1(k q)}{k q}\right]\ .
\eea
In computing $X$ and $Y$ we introduced a cut-off as we are only interested in summing modes in the linear regime. $X$ is particularly sensitive to that cut-off because  the integrand in (\ref{eq:X}) contains a term not proportional to Bessel functions that contributes  for $k\, q\gg 1$, even for $q\sim 100$ Mpc$/h$.

\begin{figure}
\begin{center}
\includegraphics[width=11.cm]{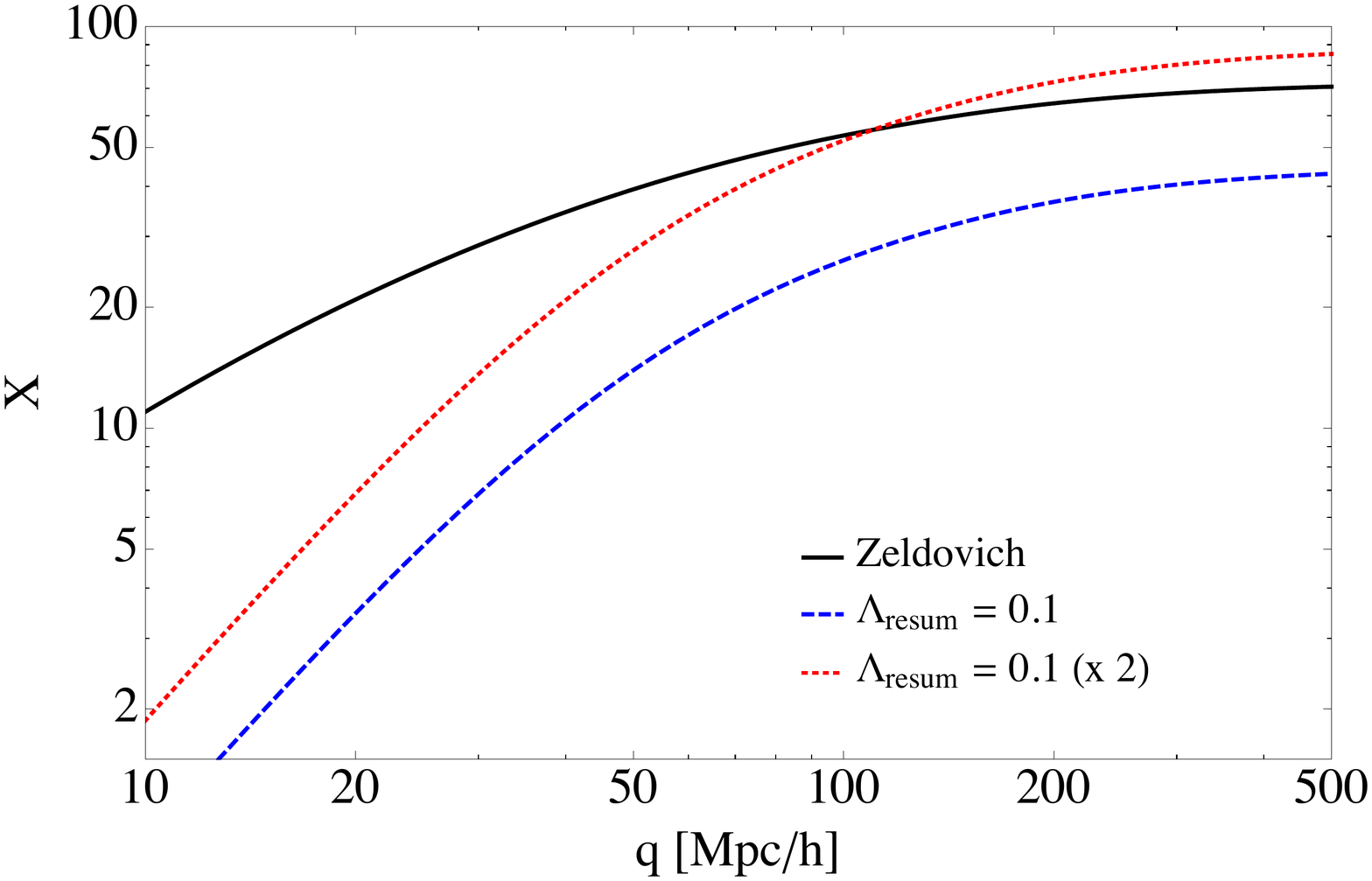}
\includegraphics[width=12.7cm]{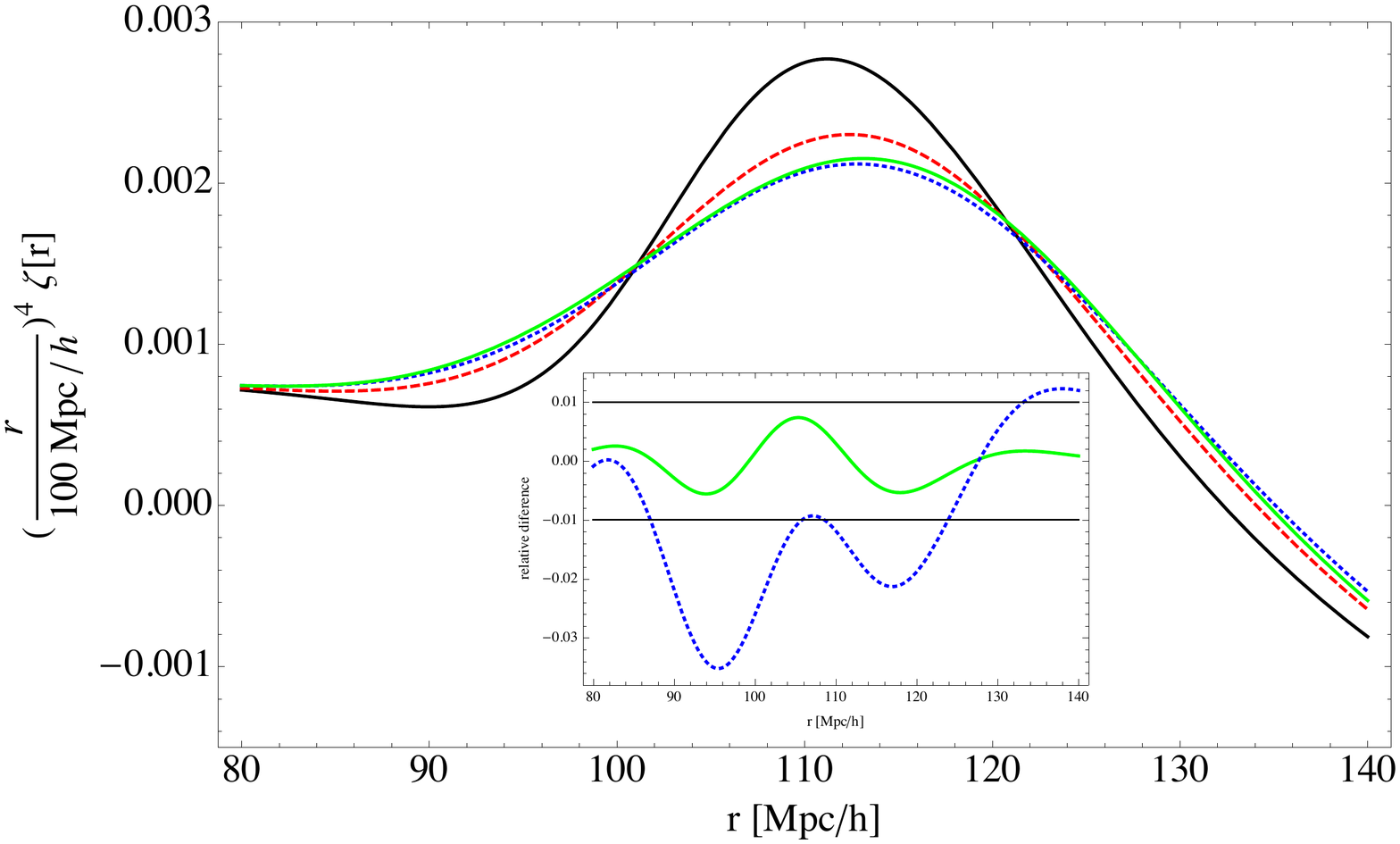}
\caption{\footnotesize  The top panel shows the linear theory value of $X$ (labeled Zeldovich) together with the result when the linear power spectrum is cut-off at $\Lambda_{\rm resum}=0.1\invMpc$. We also show the $\Lambda_{\rm resum}=0.1\invMpc$ result rescaled by a factor of two showing that this factor makes it match the Zeldovich calculation around the BAO scale. The bottom panel shows the linear theory correlation function (solid black), the zero-loop result smoothed with the original kernel (dashed red) and the rescaled one (dotted blue). We also show the two-loop result smoothed with the original kernel (solid green). In the inset we show the relative difference between the two loop calculations with the two different smoothing values (solid green) and the relative difference between the zero-loop results with the rescaled smoothing and our best answer. }
\label{irfig}
\end{center}
\end{figure}

As a result of the cut-off the value of $X$ at separations comparable to the BAO scale, which is the most relevant scale we are trying to improve with our resummation, could be substantially different from the true value of $X$. We illustrate this point in Figure \ref{irfig}. The top panel shows $X$ calculated in linear theory with no cut-off and with our $\Lambda_{\rm resum}=0.1\invMpc$. The value is significantly lower which explains why our zero-loop calculation did worse than the Zeldovich approximation. It is important to stress however that this difference is pushed to higher and higher order as we do higher loop calculations. 

To understand how sensitive our calculations are to the details of the assumptions made when computing $K_0$ we will compare our standard results with those that are obtained when we change $X$ substantially. In particular we will rescale $X$ by a factor of 2. The first question to ask is how different our two-loop answer is when we do the resummation with those two different values of $X$. The relative difference between those two cases is shown in the inset of the bottom panel of Figure \ref{irfig}. The results agree to a fraction of a percent even though the values of $X$ differed by a factor of 2 at the BAO scale.  

The freedom to adjust $X$ influences how fast we converge to the final answer, at least as far as the IR-smoothing effects are concerned. In fact we chose the factor of two rescaling to illustrate this point. With this choice the value of $X$ at the BAO scale agrees with the linear theory result with no cut-off. Roughly speaking this is the smoothing that included in the Zeldovich approximation. Indeed, in  the bottom panel of Figure \ref{irfig} we show our zero-loop result now smoothed with the rescaled value of $X$  showing that it is quite close to the final two-loop answer, within approximately 3\,\%. Clearly, as it is well known, the leading effect modifying the shape of the BAO peak is the IR smoothing. Getting that right takes you within a few percent of the answer. Doing higher loop calculations just makes our answer insensitive to the details of the IR-resummation and better calculates the dynamical effects at the non-linear scale which are needed to obtain answers accurate to a fraction of a percent. 

\section{Conclusions}\label{CONC}

We discussed how the fact that the linear power spectrum in $\Lambda$CDM is not a power law affects the relative size of the various terms in the EFT of LSS. 
We proposed a new technique to resum the effects from the long wavelength displacements that improves results for correlators computed using the Eulerian version of the EFT of LSS. We showed that our technique significantly improves earlier results by eliminating the oscillatory residuals associated with the BAO peaks that were present before and by extending the reach in the UV for quantities that are not IR safe.  

We computed the correlation function around the BAO peak and performed several internal consistency checks. We analyzed its dependence on the loop order, the EFT parameters and the details of the IR resummation technique. In future work we will compare the results directly to measurements from N-body simulations.

\subsubsection*{Acknowledgments}

We acknowledge useful discussions with  Tobias Baldauf, J.J.~Carrasco, Simon Foreman, Eiichiro Komatsu, Uros Seljak, Ravi Sheth, Rashid Sunyaev and Zvonimir Vlah.  L.S. is supported by DOE Early Career Award DE-FG02-12ER41854 and by NSF grant PHY-1068380. M.Z. is supported in part by the NSF grants  AST-0907969 and PHY-1213563. 

\begin{appendix}

\section{Computing the Smoothing Terms \label{app:smoothing-terms}}

\subsection{Computing $K_0(\vec k,\vec q;t)$}
In the main text, we have seen that given $\Sigma_0$, we can effectively resum the IR contribution by using the results of the Eulerian perturbation theory. Let us start explaining how to compute in details these terms, starting from $K_0$. Let us repeat some of the relevant formulas just for convenience. Here we follow the conventions of~\cite{Carlson:2012bu}. For simplicity, we focus on equal-time correlators, since these are the terms we compare with the data in this paper, even though the relevant formulas for unequal time correlators can be tediously but straightforwardly recovered. We therefore drop to mention the explicit time-dependence when convenient. We have 
\be
 K_0(\vec k,\vec q;t) =\exp\left[ -\frac{1}{2} \langle X_0(\vec k,\vec q;t)^2\rangle\right]\ ,
\ee
where
\be
X_0(\vec k_1,\vec q;t) =  \vec k_1 \cdot \vec{\Delta}_0(\vec q;t) \ ,
\ee 
and
\be
\vec{\Delta}_0(\vec q;t) = \vec s(\vec q_1,t)_1- \vec s(\vec q_2,t)_1\ =\int \frac{d^3 k'}{(2\pi)^3}\left(e^{i \vec k'\cdot \vec q_1}-e^{i \vec k'\cdot \vec q_2}\right)s(\vec k',t)_1\ ,
\ee
where $\vec q=\vec q_2-\vec q_1$ and where
\be
\langle s^i(\vec p_1,t)_1s^j(\vec p_2,t)_1\rangle=(2\pi)^3\delta^{(3)}(\vec p_1+\vec p_2)\; \frac{p_1^i p_1^j}{p_1^4}\; P_{\delta\delta,1}(p_1;t)\ .
\ee
After taking the expectation value, we have
\be
K_0(\vec k,\vec q;t) = \exp\left[-\frac{1}{2}A_{ij,1}(\vec q;t) k^i k^j\right] \ ,
\ee
where, by rotational invariance, we must have
\be
A_{ij,\, 1}(\vec q;t)=X(q;t)_1\, \delta_{ij}+ Y(q;t)_1\,\hat q_i \hat q_j\ .
\ee
Straightforward algebra leads to the following expressions
\bea
&& X(q;t)_1=\frac{1}{2\pi^2}\int_0^{+\infty} dk\;  \exp\left[- \frac{k^2}{\Lambda_{\rm Resum}^2}\right]\;P_{\delta\delta,11}(k;t) \left[\frac{2}{3}-2\,\frac{j_1(k q)}{k q}\right]\ , \\
&& Y(q;t)_1=\frac{1}{2\pi^2}\int_0^{+\infty} dk\;\exp\left[- \frac{k^2}{\Lambda_{\rm Resum}^2}\right]\; P_{\delta\delta,11}(k;t) \left[- 2\, j_0(k q)+6\,\frac{j_1(k q)}{k q}\right]\ ,
\eea
where $j_i(x)$ is the spherical Bessel function of kind $i$, and $\Lambda_{\rm Resum}$ represents the IR scale up to which we wish to resum the IR modes.
As for the case of the momentum correlation function, the dependence on $\Lambda_{\rm Resum}$ is supposed to represent the dependence on the remaining displacements that have not been resummed. The parameter $\epsilon_{s<}$ has been transformed into $\tilde \epsilon_{s<}\ll1\lesssim \epsilon_{s<}$. Therefore, the dependence on $\Lambda_{\rm Resum}$ is supposed to become vanishingly small as we move to higher orders in perturbation theory.  The results for the matter power spectrum presented in this paper and the dashed blue lines in Figure~\ref{dPidPi} for the momentum power spectrum and matter-momentum cross power are obtained computing $X_0$ in this way, with $\Lambda_{\rm Resum}=0.1\invMpc$. We call this procedure the non-optimized IR-resummation.

There is a trick we can perform to actually make the convergence on $\Lambda_{\rm Resum}$ even quicker. Before presenting it, we stress that this is a trick which is not parametrically justified. Implementing or not implementing the following trick should have no consequences on the ultimate result, even though, as we say, it makes the convergence in $\tilde\epsilon_{s<}$ quicker. It is quite well known that the Zeldovich approximation gives a very good approximate to actual displacement in the $\Lambda$CDM cosmology at distances of order the BAO peak~$r\simeq 120$ Mpc. For this reason, it is tempting to perform the IR resummation using IR displacement fields that share this property. We cannot send $\Lambda_{\rm Resum}$ to large values, which are the ones used in the Zeldovich approximation, as we wish to resum only long wavelength fields.  A simple way to do this is to simply rescale $X(q;t)_1$ so that it agrees at $r=120$ Mpc with the one computed with $\Lambda_{\rm Resum}=2\invMpc$. In formulas, we have

\be
X(q;t)_1\quad\to\quad  (1+\alpha) X(q;t)_1\ ,
\ee
where $\alpha$ is chose so that
\be
 (1+\alpha) \left. X(120\, {\rm Mpc}/h;t)\right|_{\Lambda_{\rm resum}=0.1 \invMpc}=\left. X(120\, {\rm Mpc}/h;t)\right|_{\Lambda_{\rm resum}=2 \invMpc}\quad\Rightarrow\quad \alpha\simeq 1\ .
\ee
We call this procedure the optimized IR-resummation, to distinguish it from the non-optimized one where we do not rescale $X_1$. We use it to show the results of the momentum power spectrum and the momentum-matter cross correlation which, being evaluated at one-loop, benefit from having a smaller $\tilde\epsilon_{s<}$. We also use it to show some results for the real space correlation function.  
We stress that this $\alpha$ parameter is not a new fitting parameter.  We have checked that this is the only effect that performing this trick achieves.

\subsection{Computing $P_{{\rm int}||_{N-j}}(r|q;t)$}

We now proceed to compute the probability of a displacement $P_{{\rm int}||_{N-j}}(r|q;t)$. This is given by the following sequence of definitions
\be
P_{{\rm int}||_{N-j}}(r|q;t)=2\pi \int_{-1}^{1} d\mu\; P_{||_{N-j}}(\vec r|\vec q;t)\ ,
\ee
where
\be
P_{||_{N-j}}(\vec r|\vec q;t)=\int\frac{d^3 k}{(2\pi)^3}\; e^{-i \vec k\cdot (\vec q-\vec r)}\;F_{||_{N-j}}(\vec q,\vec k;t)\ ,
\ee
and
\be
F_{||_{N-j}}(\vec q,\vec k;t_1,t_2)=K_0(\vec k,\vec q;t_1,t_2) \cdot\left.\left.K_0^{-1}(\vec k,\vec q;t_1,t_2)\right|\right|_{N-j} \ .
\ee
The $k$-integral to obtain  $P_{||_{N-j}}(\vec r|\vec q;t)$ is Gaussian, and can be done analytically. For $P_{||_{0,1,2}}$ that we use in this paper, it gives
\bea
&&P_{||_0}(\vec r|\vec q;t)=\frac{1}{(2\pi)^{3/2}}\frac{1}{|{A_{1}}(\vec q)|^{1/2}} e^{-\frac{1}{2}  (\vec q-\vec r)^i [{A_{1}}^{-1}]_{ij}(\vec q) (\vec q-\vec r)^j}\ , \\ \nonumber
&&P_{||_1}(\vec r|\vec q;t)=\frac{1}{(2\pi)^{3/2}}\frac{1}{|{A_{1}}(\vec q)|^{1/2}} e^{-\frac{1}{2}   (\vec q-\vec r)^i [{A_{1}}^{-1}]_{ij}(\vec q) (\vec q-\vec r)^j}\left[\frac{5}{2}-\frac{1}{2} (\vec q-\vec r)^i [{A_{1}}^{-1}]_{ij}(\vec q) (\vec q-\vec r)^j\right]\ , \\ \nonumber
&&P_{||_2}(\vec r|\vec q;t)=\ \frac{1}{(2\pi)^{3/2}}\frac{1}{|{A_{1}}(\vec q)|^{1/2}} e^{-\frac{1}{2}   (\vec q-\vec r)^i [{A_{1}}^{-1}]_{ij}(\vec q) (\vec q-\vec r)^j}\\ \nonumber
&&\qquad\qquad\qquad\times \left[\frac{35}{8}-\frac{1}{8} (\vec q-\vec r)^i [{A_{1}}^{-1}]_{ij}(\vec q) (\vec q-\vec r)^j+\frac{7}{4} \left[(\vec q-\vec r)^i [{A_{1}}^{-1}]_{ij}(\vec q) (\vec q-\vec r)^j\right]^2\right]\ ,
\eea
where, very explicitly,
\be
(\vec q-\vec r)^i [{A_{1}}^{-1}]_{ij}(\vec q) (\vec q-\vec r)^j=\frac{1}{X(q;t)_1} (q^2 + r^2 - 2 r \cdot q ) -\frac{Y(q;t)_1}{X(q;t)_1(X(q;t)_1+Y(q;t)_1)} (q - \vec r\cdot \hat q)^2\ ,
\ee
and $|{A_{1}}(\vec q)|$ is the determinant of ${A_{1}}(\vec q)$.
To obtain $P_{{\rm int}||_{N-j}}(r|q;t_1,t_2)$, one then simply integrates in the angles between $\vec q$ and $\vec r$. A software like Mathematica can do this analytically~\footnote{In evaluating the resulting functions with Mathematica for high values of the arguments, is is advisable to check that the numerical evaluation is performed correctly.}.

Finally, the matrixes $M_{||_{N-j}}(k,k',t)$ are obtained as the three dimensional Fourier transforms of $P_{{\rm int}||_{N-j}}(r|q;t)$. The  integrals over the angles between $\vec r$ and $\vec k$ and between $\vec k'$ and $\vec q$ can be done analytically, leaving us to do numerically two one-dimensional spherical  Fourier transforms. These can be easily done using fast Fourier transform (FFT), even though, since both the $r$ and the $k$ range we are interested to compute span a few orders of magnitude, it is convenient to implement the FFT in logarithmic space, following for example~\cite{Hamilton:1999uv}.

\section{\label{app:momentum_cutoff} $\bar\Lambda_{\rm resum}$ dependence of the momentum power spectrum}

In this appendix we discuss an approximate way to determine the best choice for the cutoff $\bar\Lambda_{\rm resum}(k)$ that appears in~(\ref{eq:momentum_final}). For each external $k$, we have the interest to take  $\bar\Lambda_{\rm resum}(k)$ as large as possible, so that most the of the IR contributions are resummed, but however not too high so that non-linear corrections and counterterms need not to be included. There is actually quite a simply way to determine the best choose of  $\bar\Lambda_{\rm resum}(k)$. We stress that different choices of  $\bar\Lambda_{\rm resum}(k)$ will simply differ by how much at each order the IR-effects have been resummed. The differences in the predictions obtained using the different choices will become smaller and smaller as one goes to higher orders in perturbation theory.

\begin{figure}
\begin{center}
\includegraphics[width=8.1cm]{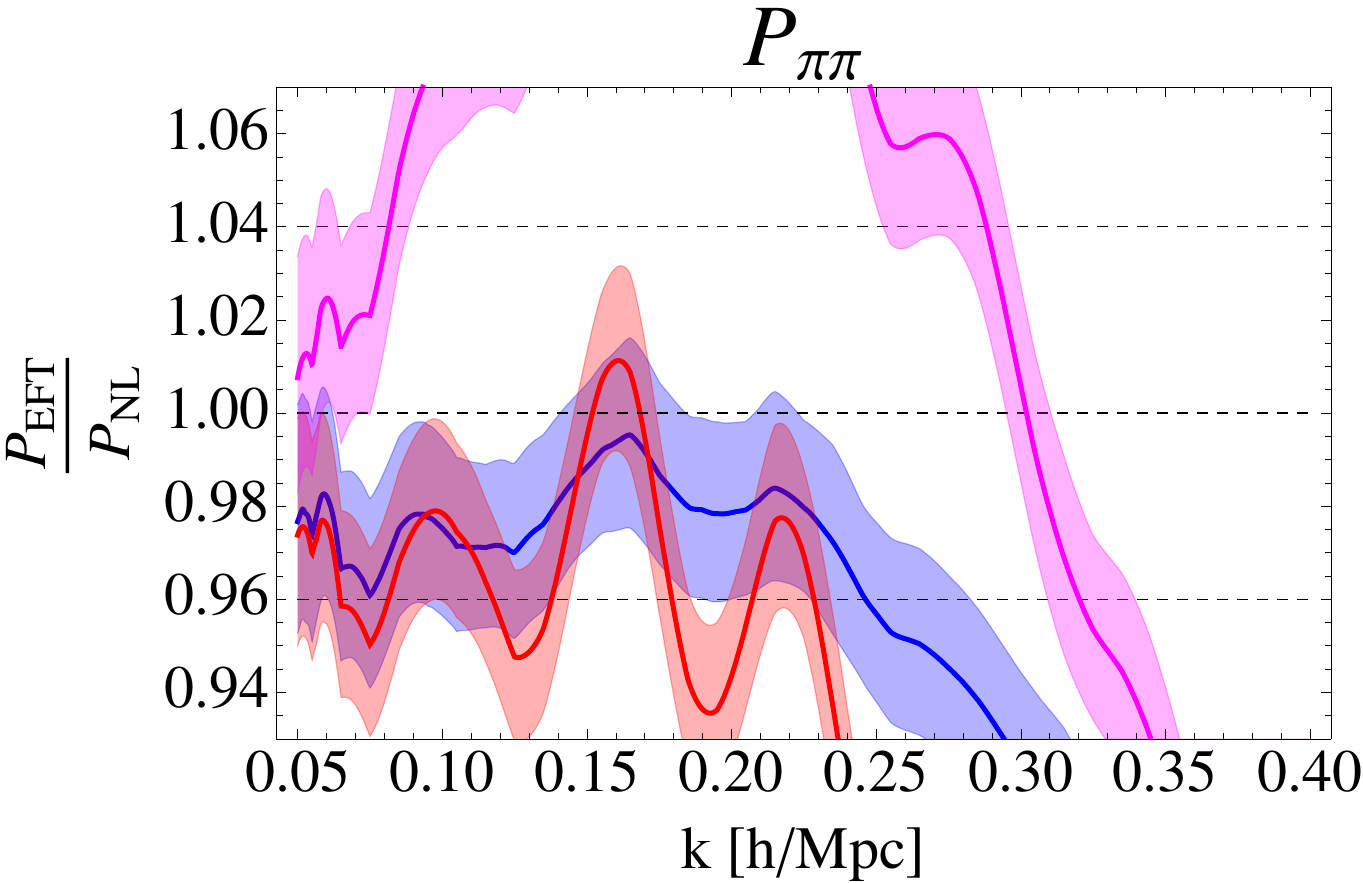}
\includegraphics[width=8.1cm]{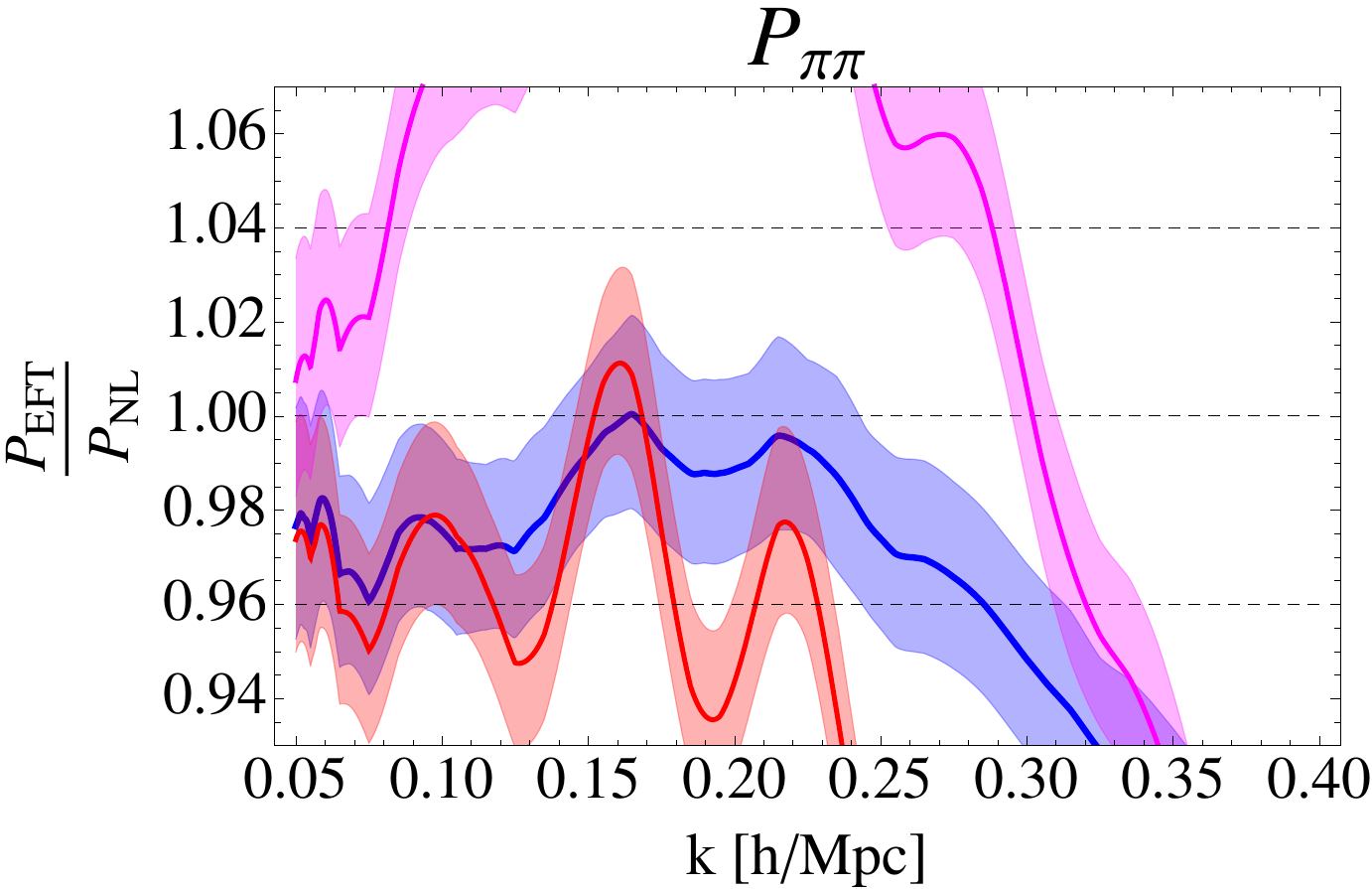}
\includegraphics[width=8.1cm]{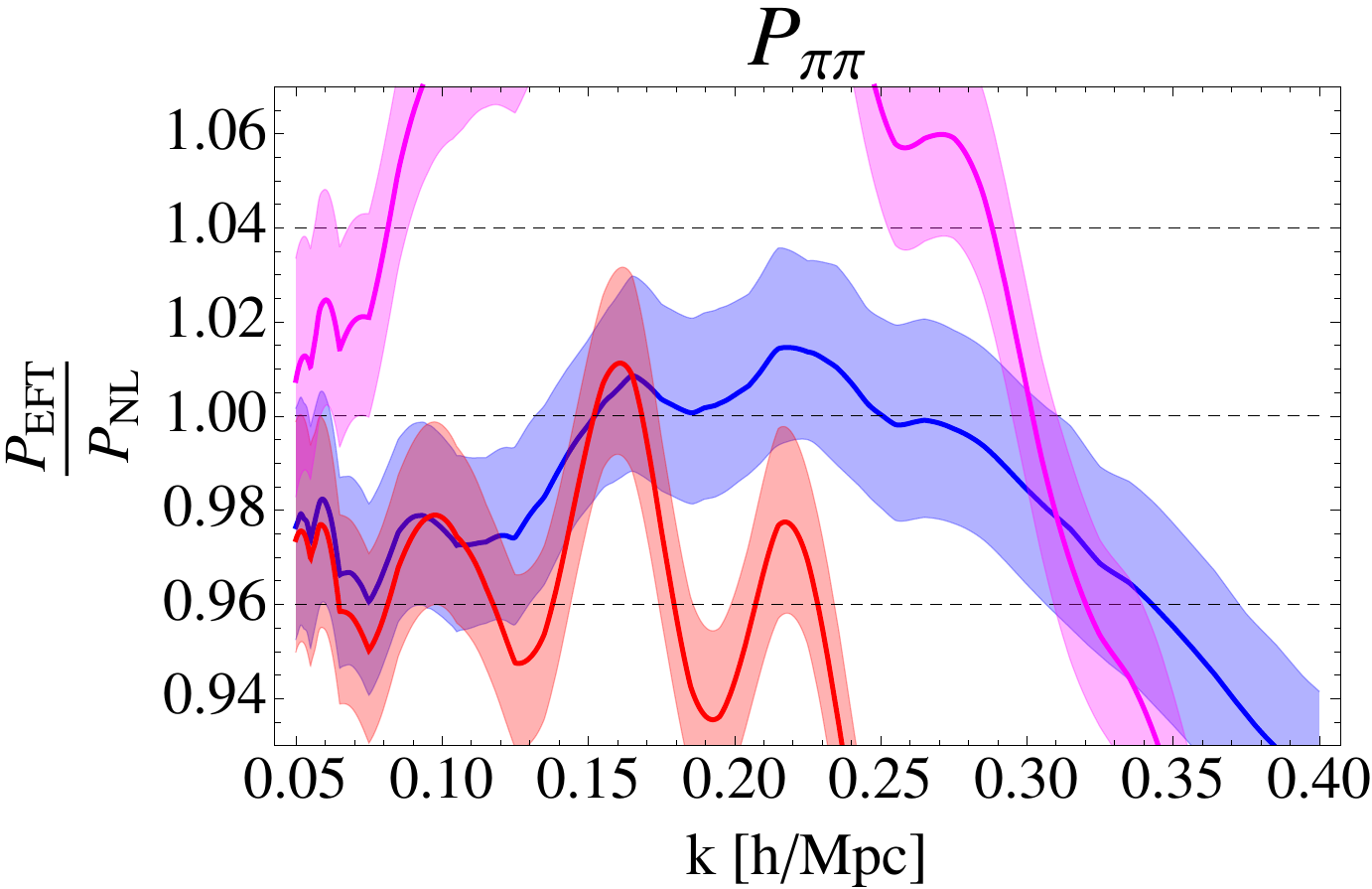}
\includegraphics[width=8.1cm]{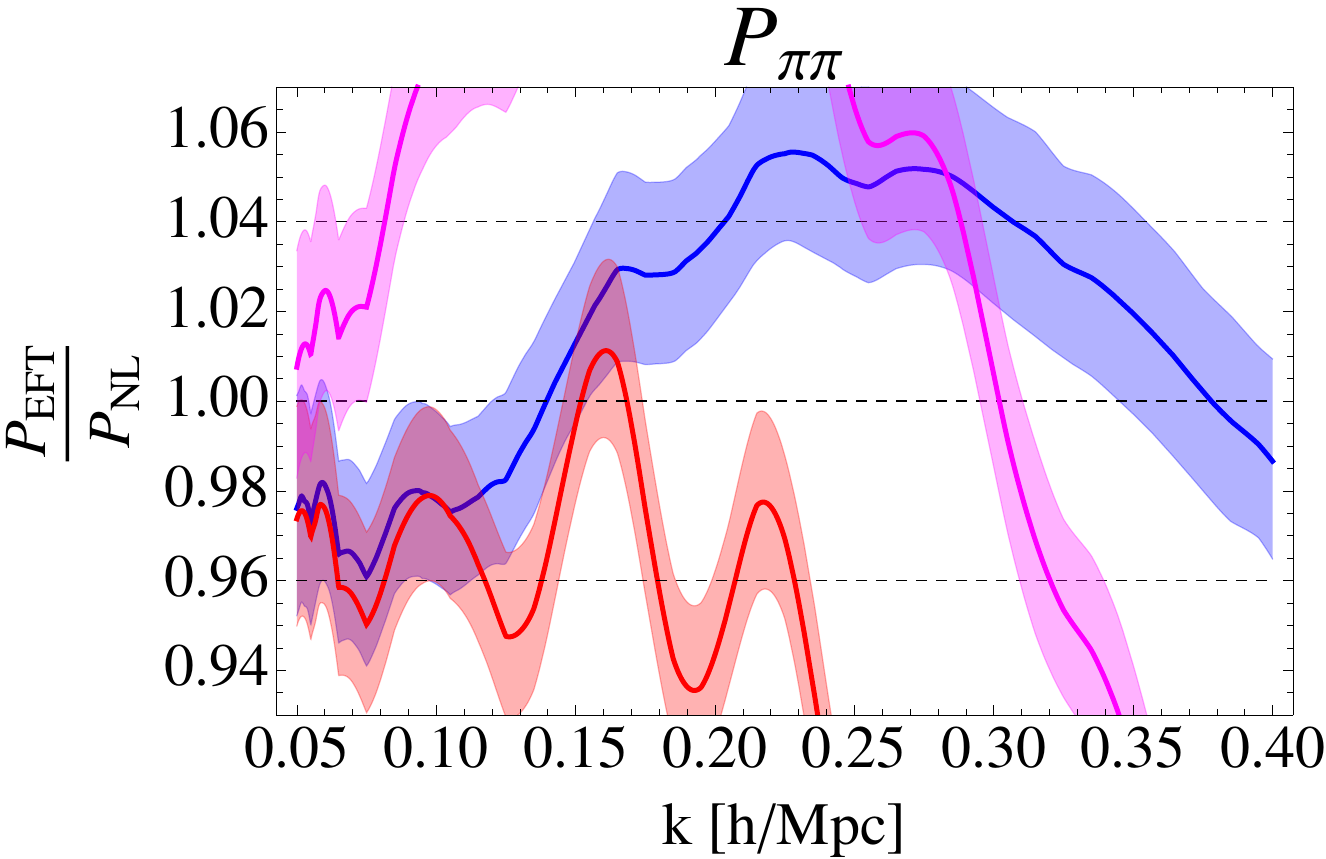}
\caption{\label{fig:momentum-different-lambdas} \footnotesize  Clockwise, the predictions of the EFT for $P_{\pi\pi}$ for $\bar\Lambda_{\rm resum}(k)=k/8,\;k/6,\;k/4$ and $k/2$. In Magenta we have the one loop SPT, in red the one-loop Eulerian EFT, and in blue we have the IR-resummed EFT. The band around each line represents the $1$-$\sigma$ cosmic variance of the simulations. One sees that the choice that best resums the IR effects is $\bar\Lambda_{\rm resum}(k)=k/6$.}
\end{center}
\end{figure}

Here in Figure~\ref{fig:momentum-different-lambdas} we plot the predictions of the momentum power spectrum using $\bar\Lambda_{\rm resum}(k)=k/8,\;k/6,\;k/4$ and $k/2$. The choice we make is $\bar\Lambda_{\rm resum}(k)=k/6$. It is pretty clear that with $\bar\Lambda_{\rm resum}(k)=k/8$, not enough of the IR modes have been resummed. This can be checked by noticing that the difference in the UV reach between the Eulerian EFT and the Lagrangian EFT is not appreciable, and for sure not close to where we expect it to be given the reach in the matter power spectrum. This means that the momentum power spectrum is still affected by IR-divergences.  $\bar\Lambda_{\rm resum}(k)=k/6$ seems fine, while when we pass to $\bar\Lambda_{\rm resum}(k)=k/4$ and $\bar\Lambda_{\rm resum}(k)=k/2$, we see that the prediction begins to  dangerously bend upward, even at relatively low $k$'s, where the slope of the Eulelrian EFT and the Langrian EFT should be quite similar. Such a mismatch could be compensated by changing the value of $\co$, but this indeed signals that with our resummation we are introducting some spurious UV terms. There is clearly some uncertainties in the determination of $\bar\Lambda_{\rm resum}(k)$, which however can be made smaller with better numerical data or with a two-loop calculation, where there is longer leverage in $k$ to check for the slope of the predicted curve. We stress that, because this procedure relies on comparing the predictions of the Eulerian and Lagrangian EFT's, the choice of $\bar\Lambda_{\rm resum}(k)$ does not correspond to a new fitting parameter.

\end{appendix}

 \begingroup\raggedright\endgroup

\end{document}